\documentclass{aa}  
\usepackage{graphicx}
\usepackage{color}
\usepackage{natbib}
\usepackage{url}
\bibpunct{(}{)}{;}{a}{}{,}
\usepackage{txfonts}
\newcounter{Rco}
\newcommand{\Ionst}[1]{\setcounter{Rco}{#1}\Roman{Rco}}
\newcommand{\Ion}[2]{\mbox{#1\,{\scriptsize\Ionst{#2}}}}
\newcommand{\Ionw}[3]{\mbox{#1\,{\scriptsize\Ionst{#2}}~$\lambda\,#3$\,\AA}}

\newcommand{\Ionww}[3]{\mbox{#1\,{\scriptsize\Ionst{#2}}~$\lambda\lambda\,#3$\,\AA}}

\newcommand{\logg}{\mbox{$\log g$}}
\newcommand{\loggw}[1]{\mbox{$\log g\hspace{-0.5mm} =\hspace{-0.5mm}  #1$}}

\newcommand{\ab}[1]{\mbox{Fig.\,\ref{#1}}}
\newcommand{\sA}[1]{\mbox{(Fig.\,\ref{#1})}}

\newcommand{\se}[1]{\mbox{Sect.\,\ref{#1}}}
\newcommand{\sga}{\raisebox{-0.10em}{$\stackrel{>}{{\mbox{\tiny $\sim$}}}$}}

\newcommand{\sla}{\raisebox{-0.10em}{$\stackrel{<}{{\mbox{\tiny $\sim$}}}$}}

\newcommand{\Teff}{\mbox{$T_\mathrm{eff}$}}
\newcommand{\Teffw}[1]{\mbox{$\Teff\hspace{-0.5mm} =\hspace{-0.5mm} #1 \,\mathrm{kK}$}}
\newcommand{\ebv}{\mbox{$E_\mathrm{B-V}$}}
\newcommand{\ebvw}[1]{\mbox{$\ebv\hspace{-0.5mm} =\hspace{-0.5mm} #1$}}
\newcommand{\nh}{\mbox{$n_\mathrm{H\,\sc{I}}$}}
\newcommand{\nhw}[1]{\mbox{$\nh\hspace{-0.5mm} =\hspace{-0.5mm} #1\, \mathrm{cm}^{-2}$}}
\newcommand{\vrad}{\mbox{$v_\mathrm{rad}$}}
\newcommand{\vradw}[1]{\mbox{$\vrad = \hspace{-0.5mm} #1\, \mathrm{km\,sec}^{-1}$}}

\newcommand{\bd}{\object{BD$-22\degr 3467$}}
\begin{document}
\title{\bd, a DAO-type star exciting the nebula \object{Abell\,35}
        \thanks
        {Based on observations with the NASA/ESA Hubble Space Telescope, obtained at the Space Telescope Science 
         Institute, which is operated by the Association of Universities for Research in Astronomy, Inc., under 
         NASA contract NAS5-26666.
        }
        \thanks
        {Based on observations made with the NASA-CNES-CSA Far Ultraviolet Spectroscopic Explorer.
        }
      }

\author{M\@. Ziegler\inst{1} 
        \and 
        T\@. Rauch\inst{1} 
        \and 
        K\@. Werner\inst{1} 
        \and
        J\@. K\"oppen\inst{2}
        \and
        J\@. W\@. Kruk\inst{3}
        }
 
\institute{Institute for Astronomy and Astrophysics,
           Kepler Center for Astro and Particle Physics,
           Eberhard Karls University, 
           Sand 1,
           72076 T\"ubingen, 
           Germany,\\
           \email{rauch@astro.uni-tuebingen.de}
           \and       
           Observatoire Astronomique de Strasbourg, 
           Universit\'e de Strasbourg,  
           11 rue de l'Universit\'e, 67000 Strasbourg, France
           \and       
           NASA Goddard Space Flight Center, Greenbelt, MD\,20771, USA}

\date{Received 4 May 2012; accepted October 12 2012}

\abstract {Spectral analyses of hot, compact stars with 
           NLTE (non-local thermodynamical equilibrium) model-atmosphere
           techniques allow the precise determination of photospheric parameters such as
           the effective temperature (\Teff), 
           the surface gravity (\logg), and  
           the chemical composition. 
           The derived photospheric metal abundances are crucial constraints for
           stellar evolutionary theory.
          }
          {Previous spectral analyses of the exciting star of the
           nebula \object{A\,35}, \bd, were based on He+C+N+O+Si+Fe models only. 
           For our analysis, we use state-of-the-art fully metal-line blanketed
           NLTE model atmospheres that consider opacities of 23 elements from hydrogen to nickel. 
           We aim to identify all observed lines in the ultraviolet (UV) spectrum of \bd\ 
           and to determine the abundances of the respective species precisely.
          }
          {For the analysis of
           high-resolution and 
           high-S/N (signal-to-noise)
           FUV (far ultraviolet, \emph{FUSE}) and 
           UV (HST/\emph{STIS}) observations, 
           we combined 
           stellar-atmosphere models and 
           interstellar line-absorption models 
           to fully reproduce the entire observed UV spectrum.
          }
          {
           The best agreement with the UV observation of \bd\ is achieved 
           at \Teffw{80\pm 10}\ and \loggw{7.2\pm 0.3}.
           While \Teff\ of previous analyses is verified,
           \logg\ is significantly lower.
           We re-analyzed lines of silicon and iron
           (1/100 and about solar abundances, respectively)
           and for the first time in this star 
           identified argon, chromium, manganese, cobalt, and nickel
           and determined abundances of
           12, 70, 35, 150, and 5 times solar, respectively.
           Our results partially agree with predictions of diffusion
           models for DA-type white dwarfs.
           A combination of 
           photospheric and interstellar line-absorption models reproduces
           more than 90\% of the observed absorption features.            
           The stellar mass is $M\,\approx\,0.48\,M_\odot$.
          }
          {
           \bd\ may not have been massive enough to ascend the 
           asymptotic giant branch and may have
           evolved directly from the extended horizontal branch to the 
           white dwarf state. This would explain why it is not surrounded
           by a planetary nebula. However, the star, ionizes the
           ambient interstellar matter, mimicking a
           planetary nebula.
          }
         

\maketitle

\color{black}
\section{Introduction}
\label{sect:intro}

\begin{figure}[ht!]\centering 
   \resizebox{\hsize}{!}{\includegraphics[angle=180]{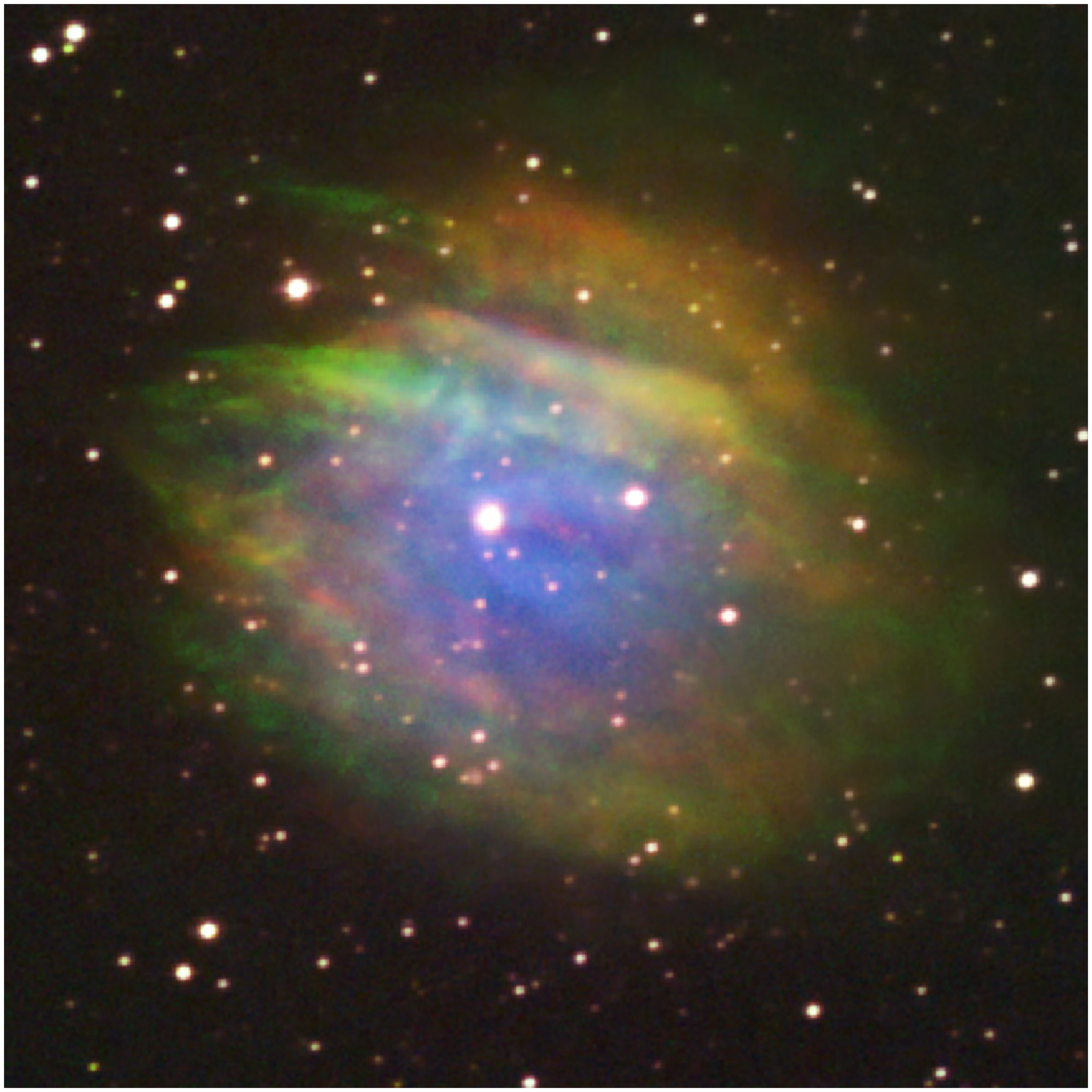}}\vspace{-9.05cm} 
   \resizebox{\hsize}{!}{\includegraphics{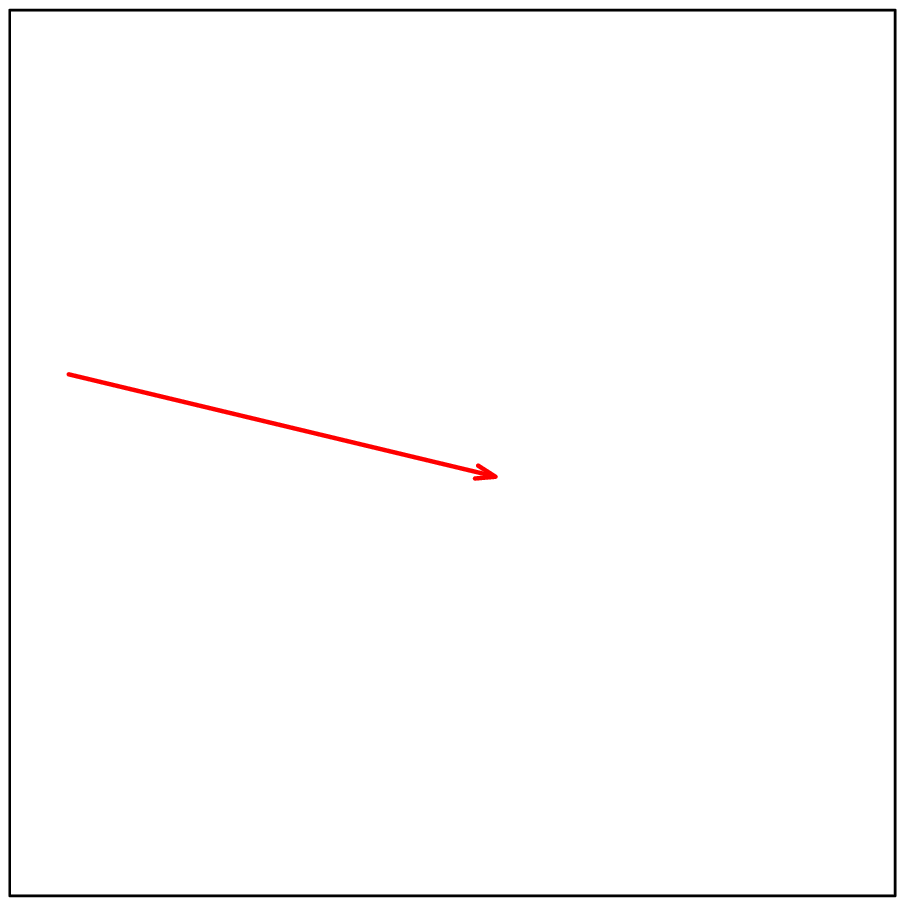}} 
    \caption{\object{A\,35}: Narrow-band composite of H$\alpha$, [\Ion{O}{3}], and [\Ion{S}{2}]
             (in total 36\,h exposure time ) by Dean Salman 
             (\object{Sh\,2$-$313}, http://www.sharplesscatalog.com/). 
              N is up, E is left, the FOV is $20\arcmin\times 20\arcmin$. 
              The (red) arrow indicates the projected motion of \bd\ during the last 10\,000 years. 
            }
   \label{fig:a35}
\end{figure}

\object{Abell\,35} 
($\alpha_\mathrm{2000}=12^\mathrm{h}53^\mathrm{m}32\fs 79$, $\delta_\mathrm{2000}=-22\degr 52\arcmin 22\farcs 55$)
was discovered by \citet[][\object{A55\,24}]{abell55} and classified as a planetary nebula (PN). 
\citet{abell66} characterized it
(\object{A66\,35}, henceforth \object{A\,35}, \object{PN\,G303.6+40.0})
as a homogeneous disk PN with an angular size between 636\arcsec\ and 938\arcsec.
Figure~\ref{fig:a35} shows an image of \object{A\,35} with very long exposure time. 
Its size is about
17\arcmin\ in east-west and 
14\arcmin\ in north-south direction.
The nebula shape appears to be quite atypical for a PN.
The bow-shock structure is surrounded by a symmetric emission area
that does not agree with the usual bi-polar or ellipsoidal morphologies.

\citet{grewing88} discovered a hot companion to the visible nucleus in the optical,
\citep[\object{SAO\,181201},][]{jacoby81}. They suggested that it is an extremely
hot DAO-type\footnote{spectral classification: characteristic \Ion{H}{1} and \Ion{He}{2} absorption lines} 
white dwarf (WD).
\object{A\,35}, \bd\ (\object{LW\,Hya}) is therefore
a resolved binary system \citep{jacoby81,demarco09}, 
composed of a 
WD star of spectral type DAO
and a G8\,III-IV\,-\,type companion \citep{unruhetal01,strassmeier09}. 
The latter dominates the spectrum at $\lambda\,\sga\,2800\,\mathrm{\AA}$ \citep{grewing88}.

\citet{borkowski90} explained the PN asymmetry
\sA{fig:a35} as the result of an interaction between the PN and the surrounding 
interstellar medium (ISM), 
the observed bow shock (the star is moving with $v = 125\,\mathrm{km/s}$)
being located at the equilibrium sphere of
ram pressure of the accelerated PN gas and the stellar wind ram pressure.

A mean (\emph{HIPPARCOS}, \emph{TYCHO-2}, and \emph{UCAC-2} values) proper motion of 
$\mu_\alpha = -60.95\pm 1.5\,\mathrm{mas/yr}$ and 
$\mu_\delta = -14.63\pm 1.4\,\mathrm{mas/yr}$
was measured by \citet{kerberetal08}
in agreement with \emph{HIPPARCOS} values \citep[HIP\,62905,][]{vanleeuven07} of
$\mu_\alpha = -60.91 \pm 1.62\,\mathrm{mas/yr}$  
$\mu_\delta = -13.41 \pm 1.24 \,\mathrm{mas/yr}$.
With its high proper motion, \bd\ passes through the complete visible nebula within
about 16\,000 years (Fig.~\ref{fig:a35}).  
Recently, \citet{frew10} claimed \object{A\,35} to be a 
``bow shock nebula in a photoionized Str\"omgren sphere in the ambient ISM''. 
This scenario does not include a PN. 
\citet{weidmanngamen11}, however, classified \object{A\,35} as a binary PN (? + G8\,IV) with a 
bc-CSPN (corresponding to binarity for the cool CSPN).

\emph{HIPPARCOS} measured a parallax of 
$7.48\pm 1.55\,\mathrm{mas}$ \citep[$D=134^{+35}_{-24}\,\mathrm{pc}$]{perryman97}.
This was corrected later by \citet{gatti98} to
$D=163^{+96}_{-58}\,\mathrm{pc}$ \citep[this value was adopted by][]{herald02}. 
\citet{gatti98} determined a separation of the hot and cool component of
$13 - 28\,\mathrm{AU}$ at this distance. However, they adopted 160\,pc as a minimum distance
because previous estimates (see their Table 2) gave much higher values
\citep[e.g\@. $360\pm 80 \,\mathrm{pc}$ from photometry measured by][]{jacoby81}.
\citet{vanleeuven07} presented a validation of the new \emph{HIPPARCOS} reduction.
The improved parallax 
($8.38\pm 1.57\,\mathrm{mas}$)
is slightly larger with a similar error range.
The corresponding distance of $D=119^{+28}_{-19}\,\mathrm{pc}$ is now even smaller.

\citet{herald02} determined the atmospheric parameters of both components
and found 
\Teffw{80}, \loggw{7.7} (cm/s$^2$) for the WD, 
and 
\Teffw{5}, \loggw{3.5} for its companion star. 
Their hot-component analysis of UV spectra 
(\emph{FUSE}\footnote{Far Ultraviolet Spectroscopic Explorer},
HST/\emph{STIS}\footnote{Hubble Space Telescope\,/\,Space Telescope Imaging Spectrograph},
and
\emph{IUE}\footnote{International Ultraviolet Explorer})
covering $905\,\mathrm{\AA}\,\sla\,\lambda\,\sla\,3280\,\mathrm{\AA}$
was performed with 
\emph{TLUSTY} and \emph{SYNSPEC}
\citep{hubeny88,hubenylanz92,hubenylanz95,hubenyetal94}.
They determined abundances for He, C, N, O, Si, and Fe (Tab.~\ref{tab:results}).
For the analysis of the cool companion they used LTE 
(local thermodynamical equilibrium) models \citep{kurucz91}. 

State-of-the-art NLTE (Non-Local Thermodynamical Equilibrium)
model atmospheres are fully metal-line blanketed
and can consider all elements from hydrogen to nickel \citep{rauch03, rauch07}.
Therefore, we decided to re-analyze the available UV spectra of \bd\ 
to determine abundances of hitherto neglected metals. 
The resulting abundance pattern can give clues to the evolutionary
history of the DAO star.
We begin with a brief description of the observed UV spectra and preparatory work (Sect.~\ref{sect:obs}), 
followed by an introduction to our model atmospheres (Sect.~\ref{sect:model}).
Then we describe our spectral analysis (Sect.~\ref{sect:ana}) in detail
and discuss the results in Sect.~\ref{sect:discussion}.

\section{Observations and interstellar absorption}
\label{sect:obs}

Hot post-AGB (asymptotic giant branch)
stars have their flux maximum in the UV wavelength range.
Most of the exhibited metal lines are located there. Metal lines of 
successive ionization stages allow one to evaluate the ionization equilibrium 
of the respective species. This is a very sensitive indicator for \Teff.
The error ranges of \Teff\ and abundance determinations can be reduced
by the analysis of many spectral lines.

The dominant ionization stages of the iron-group elements in the
relevant \Teff\ and \logg\ regime are {\sc v} and {\sc vi} (Fig.~\ref{fig:ionfrac}). 
Many strategic lines of these ions are located in the UV wavelength range. 
We therefore retrieved 
\emph{FUSE},
HST/\emph{STIS},
and 
\emph{IUE} observations 
from the MAST\footnote{\url{http://archive.stsci.edu/}} archive
(Tab.~1).

\onlfig{2}{
\begin{figure*}[ht!]\centering 
   \resizebox{0.85\hsize}{!}{\includegraphics{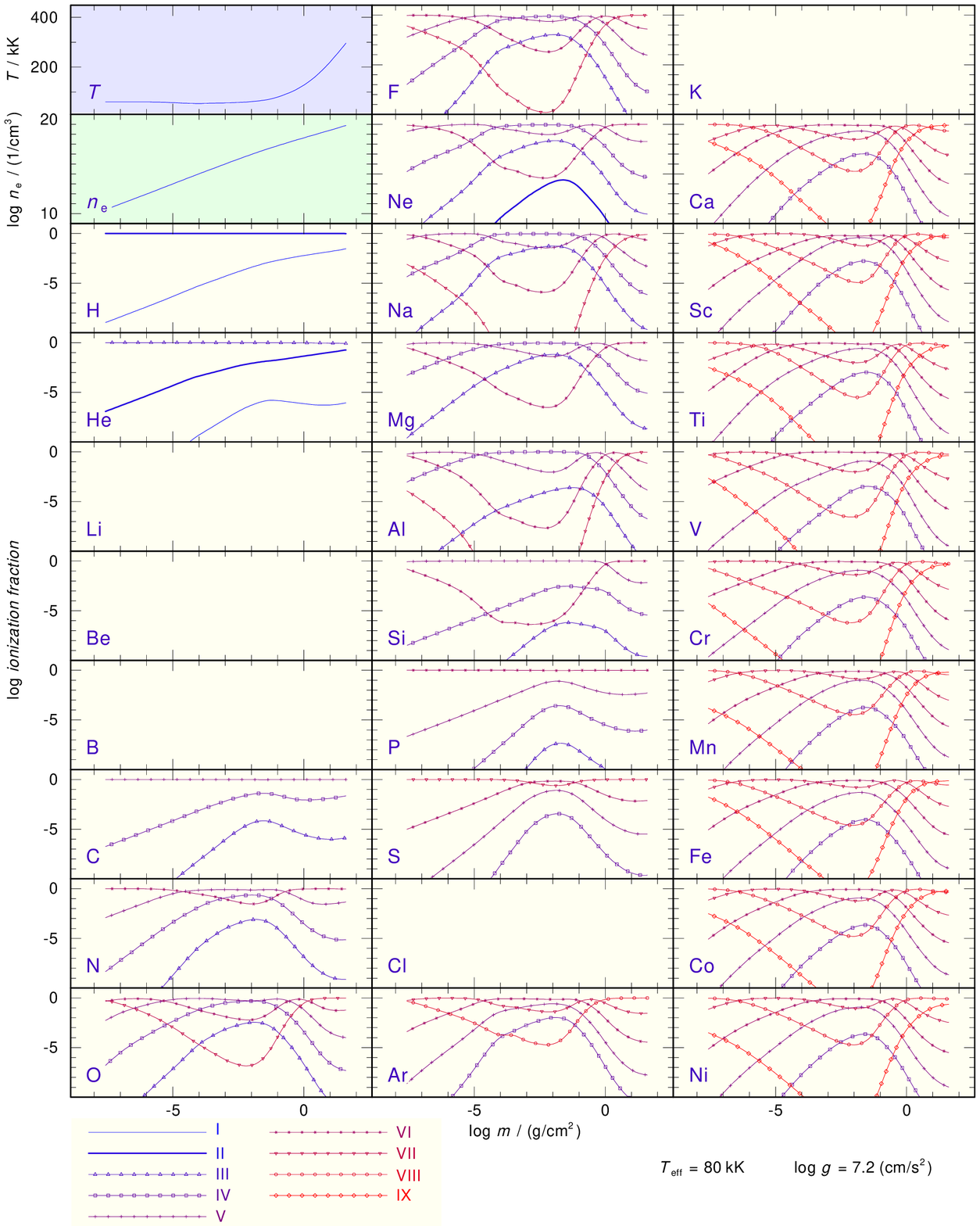}} 
    \caption{Temperature and electron density stratification
             along with ionization fractions of all elements in our final model.
            }
   \label{fig:ionfrac}
\end{figure*}
}

Our spectral analysis is based on the high-resolution 
\emph{FUSE} and 
\emph{STIS} observations.
The four \emph{STIS} spectra were co-added to improve the S/N.
\citet{herald02} analyzed the same \emph{FUSE} spectrum but only
one (O4GT02010) of the \emph{STIS} spectra.
The low-resolution \emph{IUE}\footnote{International Ultraviolet Explorer} observations were used
in addition to determine the interstellar reddening (\ebv, Fig.~\ref{fig:ebv}) as well as to
verify the flux calibration.
The \emph{STIS} observation had to be scaled by a factor of $1.995$ to match the flux level of
the \emph{IUE} observation. This offset is possibly due to bad centering of
the exciting star during the exposure \citep{herald02}.

\onltab{1}{
\begin{table}[ht!]\centering
\label{tab:obs}
\caption{Log of the UV observations of \bd.
         The \emph{FUSE} observation was performed with the LWRS 
         (low-resolution aperture,
          resolving power $R = \lambda/\Delta\lambda \approx 20\,000$), 
         \emph{STIS} with grating E140M ($R \approx 45\,800$), and
         \emph{IUE} in low-resolution mode ($R \approx 300$). 
         }         
\begin{tabular}{ccccccc}
\hline
\hline
\noalign{\smallskip}
Instrument  & ObsId        & Start time           &  Exp\@. time      \\
            &              &   (UT)               &  (s)              \\
\hline                                      
\noalign{\smallskip}                        
\emph{FUSE} &  P1330101000 & 2000-05-20 20:27:37  &  4\,416           \\
 \hline                                     
\noalign{\smallskip}                        
\emph{STIS} &  O4GT02010   & 1999-04-17 21:14:49  &  2\,050           \\
            &  O4GT02020   & 1999-04-17 22:37:03  &  2\,800           \\
            &  O4GT02030   & 1999-04-18 00:16:10  &  2\,740           \\
            &  O4GT02040   & 1999-04-18 01:52:54  &  2\,740           \\
\hline                                     
\noalign{\smallskip}                        
\emph{IUE}  &  SWP44121LL  & 1992-03-06 04:50:23  &  1\,440           \\
            &  LWP22518LL  & 1992-03-06 05:26:54  &  1\,440           \\
\hline
\end{tabular}
\end{table}  
}

\subsection{Interstellar absorption and reddening} 
\label{sect:ismabsred}

\begin{figure}[ht!] 
   \resizebox{\hsize}{!}{\includegraphics{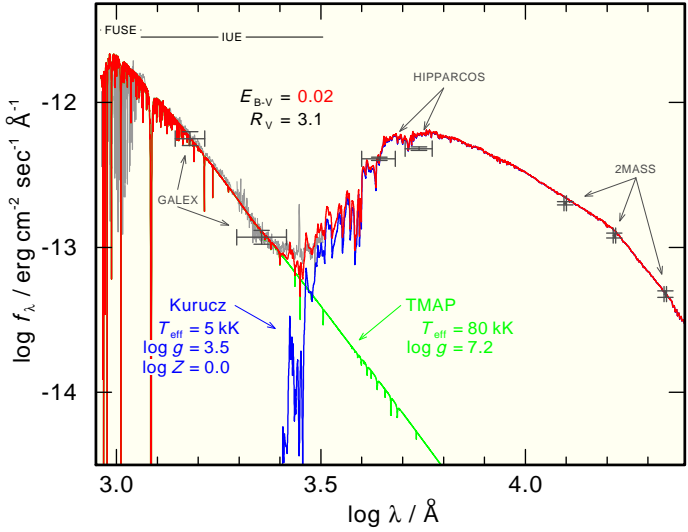}} 
    \caption{UV-optical-IR flux distribution of the \bd\ binary system,
             fitted by atmosphere models for both components.  
            }
   \label{fig:ebv}
\end{figure}

The general procedure of determining the interstellar reddening is
to normalize a theoretical spectrum as far as possible in the low-energy
range because it is not significant there.
In the case of \bd, the companion
star dominates the observation at wavelengths longer 
than $\approx 2\,800\,\AA$. Therefore, we constructed a theoretical, composite spectrum
(with appropriate stellar radii, $R^2_\mathrm{pri}/R^2_\mathrm{s} = 7.43\cdot 10^{-5}$)
using our 
final model spectrum (\Teffw{80}, \loggw{7.2}) and 
a
Kurucz model\footnote{\url{ftp://ftp.stsci.edu/cdbs/grid/k93models/kp00/}}
($T_\mathrm{eff}=5\,000\,\rm{K}$, $\loggw{3.5}$, solar abundances $\log Z = 0.0$) 
for the hot and cool component, respectively.
This spectrum was then normalized
to the 2MASS K flux (Fig.~\ref{fig:ebv}).
The best fit (matching all \emph{GALEX}, \emph{HIPPARCOS}, and \emph{2MASS} magnitudes
and the UV spectra) was achieved for \ebvw{0.02 \pm 0.02}, using the reddening law of 
\citet{fitzpatrick99} with the standard $R_\mathrm{v}=3.1$. 
Our value agrees within error limits with those of \citet[\ebvw{0.04\pm 0.01}]{herald02}, 
who used the Rayleigh-Jeans tail of the UV observations 
($1\,000\,\mathrm{\AA}\,\sla\,\lambda\,\sla\,2\,800\,\mathrm{\AA}$) only.

To determine the column density of interstellar neutral
hydrogen, \nh, models with different \nh\ were compared with
Ly\,$\alpha$. We found a best fit for \nhw{5.0\pm 1.5 \times 10^{20}} \sA{fig:nh},
which agrees with the value found by \citet{herald02}.

\begin{figure}[ht!] 
   \resizebox{\hsize}{!}{\includegraphics{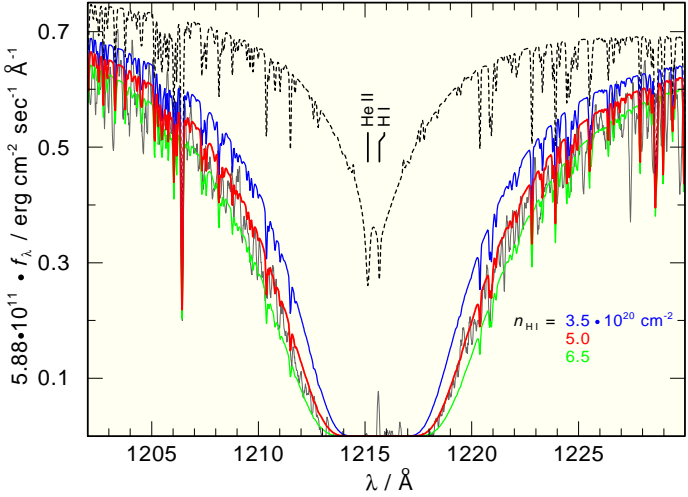}} 
    \caption{\emph{STIS} observation around \Ion{H}{1} Ly\,$\alpha$ compared with our
             final model (Tab\@.~\ref{tab:results}) at different $n_\mathrm{H\,I}$. 
             The dashed line is the WD's atmospheric flux.    
            }
   \label{fig:nh}
\end{figure}

\subsection{Radial velocity of the hot component}
\label{sect:radvel}

The \emph{STIS} observation shows numerous sharp, isolated photospheric lines 
that are suitable to determine \vrad. 
On average 
(21 lines of 
\Ion{N}{5},
\Ion{O}{4},
\Ion{O}{5},
\Ion{Ar}{5},
\Ion{Cr}{5},
\Ion{Cr}{6},
\Ion{Fe}{5},
\Ion{Fe}{6},
and \Ion{Ni}{5}), 
we measured \vradw{-14.4\pm 0.7}.

\section{Model atmospheres and atomic data}
\label{sect:model}

In this section, we briefly describe the atomic data and the programs 
used for our analysis. Details about the calculation of the
stellar atmosphere models can be found in \se{subsec:photo}. The usage of
\emph{OWENS} for the calculation of the ISM absorption
is described in \se{subsect:mod:ism}. 

\onltab{2}{
\begin{table}[ht!]\centering
\caption{Statistics of H -- Ar model atoms used in our calculations}         
\label{tab:statistics}
\begin{tabular}{lrrrlrrr}
\hline
\hline
\noalign{\smallskip}
            & \multicolumn{2}{c}{levels} & &             & \multicolumn{2}{c}{levels} & \\
\cline{2-3}
\cline{6-7}
\noalign{\smallskip}
\hbox{}\hspace{2mm}ion         & NLTE & LTE & lines  & \hbox{}\hspace{2mm}ion         & NLTE & LTE & lines   \\
\hline         
\noalign{\smallskip}
\Ion{\hbox{}\hspace{1.5mm}H}{1}  &   12 &    4 &    66 &  \Ion{\hbox{}\hspace{0.5mm}Na}{6} &   43 &   10 &   130 \\
\Ion{\hbox{}\hspace{1.5mm}H}{2}  &    1 &    0 &   $-$ &  \Ion{\hbox{}\hspace{0.5mm}Na}{7} &    1 &    0 &     0 \\
\Ion{He}{1} &    5 &   98 &     3 &  \Ion{Mg}{3} &    1 &   34 &     0 \\
\Ion{He}{2} &   16 &   16 &   120 &  \Ion{Mg}{4} &   31
 &    0 &    93 \\
\Ion{He}{3} &    1 &    0 &   $-$ &  \Ion{Mg}{5} &   15 &   37 &    18 \\
\Ion{\hbox{}\hspace{1.6mm}C}{3}  &    6 &   61 &    12 &  \Ion{Mg}{6} &    1 &    0 &     0 \\
\Ion{\hbox{}\hspace{1.6mm}C}{4}  &   54 &    4 &   295 &  \Ion{\hbox{}\hspace{1.3mm}Al}{3} &    1 &    6 &     0 \\
\Ion{\hbox{}\hspace{1.6mm}C}{5}  &    1 &    0 &     0 &  \Ion{\hbox{}\hspace{1.3mm}Al}{4} &   15 &    2 &     0 \\
\Ion{\hbox{}\hspace{1.5mm}N}{3}  &    1 &   65 &     0 &  \Ion{\hbox{}\hspace{1.3mm}Al}{5} &    1 &   16 &     0 \\
\Ion{\hbox{}\hspace{1.5mm}N}{4}  &   16 &   78 &    30 &  \Ion{\hbox{}\hspace{1.3mm}Al}{6} &   14 &   24 &    16 \\
\Ion{\hbox{}\hspace{1.5mm}N}{5}  &   54 &    8 &   297 &  \Ion{\hbox{}\hspace{1.3mm}Al}{7} &    1 &    0 &     0 \\
\Ion{\hbox{}\hspace{1.5mm}N}{6}  &    1 &    0 &     0 &  \Ion{\hbox{}\hspace{1.6mm}Si}{3} &    3 &   31 &     1 \\
\Ion{\hbox{}\hspace{1.5mm}O}{3}  &    3 &   69 &     0 &  \Ion{\hbox{}\hspace{1.6mm}Si}{4} &   16 &    7 &    44 \\
\Ion{\hbox{}\hspace{1.5mm}O}{4}  &   18 &   67 &    39 &  \Ion{\hbox{}\hspace{1.6mm}Si}{5} &   25 &    0 &    59 \\
\Ion{\hbox{}\hspace{1.5mm}O}{5}  &   90 &   36 &   610 &  \Ion{\hbox{}\hspace{1.6mm}Si}{6} &    1 &    0 &     0 \\
\Ion{\hbox{}\hspace{1.5mm}O}{6}  &   54 &    8 &   291 &  \Ion{\hbox{}\hspace{2.4mm}P}{3}  &    1 &    9 &     0 \\
\Ion{\hbox{}\hspace{1.5mm}O}{7}  &    1 &    0 &     0 &  \Ion{\hbox{}\hspace{2.4mm}P}{4}  &   15 &   36 &     9 \\
\Ion{\hbox{}\hspace{1.9mm}F}{3}  &    1 &    6 &     0 &  \Ion{\hbox{}\hspace{2.4mm}P}{5}  &   18 &    7 &    12 \\
\Ion{\hbox{}\hspace{1.9mm}F}{4}  &    1 &   10 &     0 &  \Ion{\hbox{}\hspace{2.4mm}P}{6}  &    1 &    0 &     0 \\
\Ion{\hbox{}\hspace{1.9mm}F}{5}  &   15 &   91 &    31 &  \Ion{\hbox{}\hspace{2.4mm}S}{4}  &    6 &   94 &     4 \\
\Ion{\hbox{}\hspace{1.9mm}F}{6}  &   12 &  115 &    16 &  \Ion{\hbox{}\hspace{2.4mm}S}{5}  &   21 &   89 &    37 \\
\Ion{\hbox{}\hspace{1.9mm}F}{7}  &    1 &    0 &     0 &  \Ion{\hbox{}\hspace{2.4mm}S}{6}  &   18 &   19 &    48 \\
\Ion{Ne}{2} &    1 &   33 &     0 &  \Ion{\hbox{}\hspace{2.0mm}S}{7}  &    1 &    0 &     0 \\
\Ion{Ne}{3} &    3 &   43 &     0 &  \Ion{\hbox{}\hspace{0.8mm}Ar}{4} &    1 &  349 &     0 \\
\Ion{Ne}{4} &    3 &   37 &     0 &  \Ion{\hbox{}\hspace{0.8mm}Ar}{5} &   32 &  329 &    38 \\
\Ion{Ne}{5} &   20 &   74 &    35 &  \Ion{\hbox{}\hspace{0.8mm}Ar}{6} &   16 &  168 &    21 \\
\Ion{Ne}{6} &    1 &    0 &     0 &  \Ion{\hbox{}\hspace{0.8mm}Ar}{7} &   40 &  112 &   130 \\
\Ion{Na}{3} &    1 &  186 &     0 &  \Ion{\hbox{}\hspace{0.8mm}Ar}{8} &    1 &    0 &     0 \\               
\Ion{Na}{4} &    1 &  237 &     0 &              &      &      &       \\      
\Ion{Na}{5} &    8 &    42 &    9 &  \hbox{}\hspace{2.0mm}total       &  741 & 2837 &  2514 \\
\hline
\end{tabular}
\end{table}  
}

\addtocounter{table}{-1}

\onltab{2}{
\begin{table}[ht!]\centering
\caption{Continued for Ca -- Ni.
The so-called ``sample lines'' are combined to ``super lines'' by {\sc IrOnIc}.}         
\begin{tabular}{lrrrrr}
\hline
\hline
\noalign{\smallskip}
\hbox{}\hspace{2mm}ion          & \multicolumn{2}{c}{super levels} & \multicolumn{2}{c}{super lines} & sample lines  \\
\noalign{\smallskip}
\hline         
\noalign{\smallskip}
\Ion{\hbox{}\hspace{1.2mm}Ca}{4}  &  \hbox{}\hspace{7mm}6&  & \hbox{}\hspace{7mm}16 &  &     20\,291 \\  
\Ion{\hbox{}\hspace{1.2mm}Ca}{5}  &                     6&  &                    21 &  &     141\,956 \\
\Ion{\hbox{}\hspace{1.2mm}Ca}{6}  &                     6&  &                    19 &  &     114\,545 \\
\Ion{\hbox{}\hspace{1.2mm}Ca}{7}  &                     6&  &                    21 &  &      7\,1608 \\
\Ion{\hbox{}\hspace{1.2mm}Ca}{8}  &                     6&  &                    20 &  &       9\,124 \\
\Ion{\hbox{}\hspace{1.2mm}Ca}{9}  &                     1&  &                     0 &  &            0 \\
\noalign{\smallskip}                                                                       
\Ion{\hbox{}\hspace{1.6mm}Sc}{4}  &                     6&  &                    20 &  &      15\,024 \\
\Ion{\hbox{}\hspace{1.6mm}Sc}{5}  &                     6&  &                    21 &  &     261\,235 \\
\Ion{\hbox{}\hspace{1.6mm}Sc}{6}  &                     6&  &                    19 &  &     237\,271 \\
\Ion{\hbox{}\hspace{1.6mm}Sc}{7}  &                     6&  &                    20 &  &     176\,143 \\
\Ion{\hbox{}\hspace{1.6mm}Sc}{8}  &                     6&  &                    21 &  &      91\,935 \\
\Ion{\hbox{}\hspace{1.6mm}Sc}{9}  &                     1&  &                     0 &  &            0 \\
\noalign{\smallskip}                                                                       
\Ion{\hbox{}\hspace{1.8mm}Ti}{4}  &                     6&  &                    19 &  &       1\,000 \\
\Ion{\hbox{}\hspace{1.8mm}Ti}{5}  &                     6&  &                    20 &  &      26\,654 \\
\Ion{\hbox{}\hspace{1.8mm}Ti}{6}  &                     6&  &                    19 &  &      95\,448 \\
\Ion{\hbox{}\hspace{1.8mm}Ti}{7}  &                     6&  &                    20 &  &     230\,618 \\
\Ion{\hbox{}\hspace{1.8mm}Ti}{8}  &                     6&  &                    21 &  &     182\,699 \\
\Ion{\hbox{}\hspace{1.8mm}Ti}{9}  &                     1&  &                     0 &  &            0 \\
\noalign{\smallskip}                                                                       
\Ion{\hbox{}\hspace{2.2mm}V}{4}   &                     6&  &                    19 &  &      3\,7130 \\
\Ion{\hbox{}\hspace{2.2mm}V}{5}   &                     6&  &                    20 &  &       2\,123 \\
\Ion{\hbox{}\hspace{2.2mm}V}{6}   &                     6&  &                    19 &  &      35\,251 \\
\Ion{\hbox{}\hspace{2.2mm}V}{7}   &                     6&  &                    19 &  &     112\,883 \\
\Ion{\hbox{}\hspace{2.2mm}V}{8}   &                     6&  &                    20 &  &     345\,089 \\
\Ion{\hbox{}\hspace{2.2mm}V}{9}   &                     1&  &                     0 &  &            0 \\
\noalign{\smallskip}                                                                       
\Ion{\hbox{}\hspace{1.4mm}Cr}{4}  &                     6&  &                    20 &  &     234\,170 \\
\Ion{\hbox{}\hspace{1.4mm}Cr}{5}  &                     6&  &                    20 &  &      43\,860 \\
\Ion{\hbox{}\hspace{1.4mm}Cr}{6}  &                     6&  &                    20 &  &       4\,406 \\
\Ion{\hbox{}\hspace{1.4mm}Cr}{7}  &                     6&  &                    19 &  &      37\,070 \\
\Ion{\hbox{}\hspace{1.4mm}Cr}{8}  &                     6&  &                    20 &  &     132\,221 \\
\Ion{\hbox{}\hspace{1.4mm}Cr}{9}  &                     1&  &                     0 &  &            0 \\
\noalign{\smallskip}                                                                       
\Ion{Mn}{4}  &                                          6&  &                    20 &  &     719\,387 \\
\Ion{Mn}{5}  &                                          6&  &                    20 &  &     285\,376 \\
\Ion{Mn}{6}  &                                          6&  &                    20 &  &      70\,116 \\
\Ion{Mn}{7}  &                                          6&  &                    20 &  &       8\,277 \\
\Ion{Mn}{8}  &                                          6&  &                    20 &  &      37\,168 \\
\Ion{Mn}{9}  &                                          1&  &                     0 &  &            0 \\
\noalign{\smallskip}                                                    
\Ion{\hbox{}\hspace{1.5mm}Fe}{4}  &                     6&  &                    20 &  &  3\,102\,371 \\
\Ion{\hbox{}\hspace{1.5mm}Fe}{5}  &                     6&  &                    20 &  &  3\,266\,247 \\
\Ion{\hbox{}\hspace{1.5mm}Fe}{6}  &                     6&  &                    20 &  &     991\,935 \\
\Ion{\hbox{}\hspace{1.5mm}Fe}{7}  &                     6&  &                    20 &  &     200\,455 \\
\Ion{\hbox{}\hspace{1.5mm}Fe}{8}  &                     6&  &                    18 &  &      19\,587 \\
\Ion{\hbox{}\hspace{1.5mm}Fe}{9}  &                     1&  &                     0 &  &            0 \\
\noalign{\smallskip}                                                    
\Ion{\hbox{}\hspace{1.0mm}Co}{4}  &                     6&  &                    20 &  &     552\,916 \\
\Ion{\hbox{}\hspace{1.0mm}Co}{5}  &                     6&  &                    20 &  &  1\,469\,717 \\
\Ion{\hbox{}\hspace{1.0mm}Co}{6}  &                     6&  &                    18 &  &     898\,484 \\
\Ion{\hbox{}\hspace{1.0mm}Co}{7}  &                     6&  &                    19 &  &     492\,913 \\
\Ion{\hbox{}\hspace{1.0mm}Co}{8}  &                     6&  &                    20 &  &      88\,548 \\
\Ion{\hbox{}\hspace{1.0mm}Co}{9}  &                     1&  &                     0 &  &            0 \\
\noalign{\smallskip}                                                    
\Ion{\hbox{}\hspace{1.5mm}Ni}{4}  &                     6&  &                    20 &  &  2\,512\,561 \\
\Ion{\hbox{}\hspace{1.5mm}Ni}{5}  &                     6&  &                    20 &  &  2\,766\,664 \\
\Ion{\hbox{}\hspace{1.5mm}Ni}{6}  &                     6&  &                    18 &  &  7\,408\,657 \\
\Ion{\hbox{}\hspace{1.5mm}Ni}{7}  &                     6&  &                    18 &  &  4\,195\,381 \\
\Ion{\hbox{}\hspace{1.5mm}Ni}{8}  &                     6&  &                    20 &  &  1\,473\,122 \\
\Ion{\hbox{}\hspace{1.5mm}Ni}{9}  &                     1&  &                     0 &  &            0 \\
\hline                                                                  
\noalign{\smallskip}                                                    
\hbox{}\hspace{6mm}total        &                   279&  &                     884 &  & 33\,219\,636 \\
\hline
\end{tabular}
\end{table}  
}

\subsection{The photospheric model for the hot component}
\label{subsec:photo}

Our model atmospheres 
(plane-parallel, 
chemically homogeneous,
in hydrostatic and radiative equilibrium) 
are calculated with
\emph{TMAP}\footnote{\url{http://astro.uni-tuebingen.de/~TMAP}}, 
the T\"ubingen NLTE model atmosphere package \citep{werner03}.
We calculated a small grid of models that
include opacities of 23 elements from H -- Ni (Tab.~\ref{tab:statistics}).

H -- Ar are represented by ``classical'' model atoms \citep{rauch97} taken from 
\emph{TMAD}\footnote{\url{http://astro.uni-tuebingen.de/~TMAD}}, 
the T\"ubingen model atom database. 
For Ca -- Ni, a statistical approach was applied 
to handle the large number of atomic levels and line transitions. 
We employed {\sc IrOnIc} \citep{rauchdeetjen03, rauch03} to
consider thousands of levels and millions of lines provided by \citet[and priv.~comm.]{kurucz09} 
and the Opacity Project \citep{seaton94}. 

\begin{figure*}[ht!] 
   \resizebox{\hsize}{!}{\includegraphics{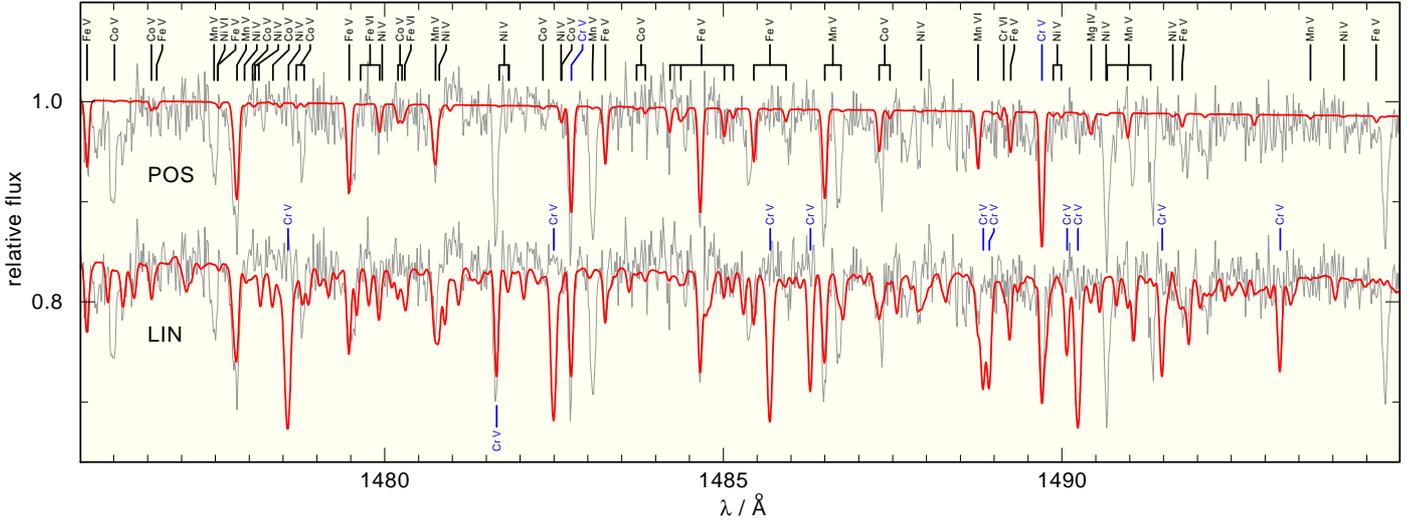}} 
    \caption{Section of the \emph{STIS} spectrum (black line) compared with
             our \emph{TMAP} model. The SED in the upper panel was calculated using Kurucz's
             POS data 
             (the strongest lines are identified at top), 
             the SED in the lower panel using LIN data. 
             This section contains mainly Fe\,{\sc v} - {\sc vi} and Mn\,{\sc v} - {\sc vi} lines.
             The strongest \Ion{Cr}{5} LIN lines are identified at the bottom.  
            }
   \label{fig:stis}
\end{figure*}

Kurucz`s line lists are divided into so-called LIN (measured and theoretical lines)
and POS (only measured, ``good'' wavelengths) lists. 
For the model-atmosphere calculations,
the LIN lists were used, to consider the total opacity properly.
To calculate the emerging spectrum, POS lists were used to identify iron-group lines. 
Figure~\ref{fig:stis} demonstrates the difference between LIN and POS line lists.
As an example for all species and their ions serves \Ion{Cr}{5}
(within the shown 1470 - 1500\,\AA\ interval).
At present, Kurucz provides 73\,222 LIN and 249 POS lines of \Ion{Cr}{5}, 459 LIN and 6 POS lines
in this wavelength interval.
We adjusted the Cr abundance to $\mbox{[Cr]} = +3.1$ to reproduce
the two strongest \Ion{Cr}{5} POS lines, \Ionww{Cr}{5}{1482.76, 1489.71}.
At this abundance, most of the prominent \Ion{Cr}{5} LIN lines appear displaced or
simply too strong. An exception is \Ionw{Cr}{5}{1481.66}, which matches the observation well.
It can not be excluded that the \Ionw{Cr}{5}{1490.24} LIN lines in the model is
the observed 1490.66\,\AA\ line, for instance. 
However, reliable abundance determinations are only possible using unambiguously identified 
POS lines (Sect.~\ref{sect:ana:metal}).

In the framework of the Virtual
Observatory (\emph{VO}\footnote{\url{http://www.ivoa.net}}), all spectral
energy distributions (SEDs, $\lambda - F_\lambda$) calculated from our
model-atmosphere grid are available in \emph{VO}-compliant form
from the registered \emph{VO} service  
\emph{TheoSSA}\footnote{\url{http://dc.g-vo.org/theossa}}
provided by the \emph{German Astrophysical Virtual Observatory}
(\emph{GAVO}\footnote{\url{http://www.g-vo.org}}).

\subsection{The ISM line-absorption model}
\label{subsect:mod:ism}

\begin{figure*}[ht!] 
   \resizebox{\hsize}{!}{\includegraphics{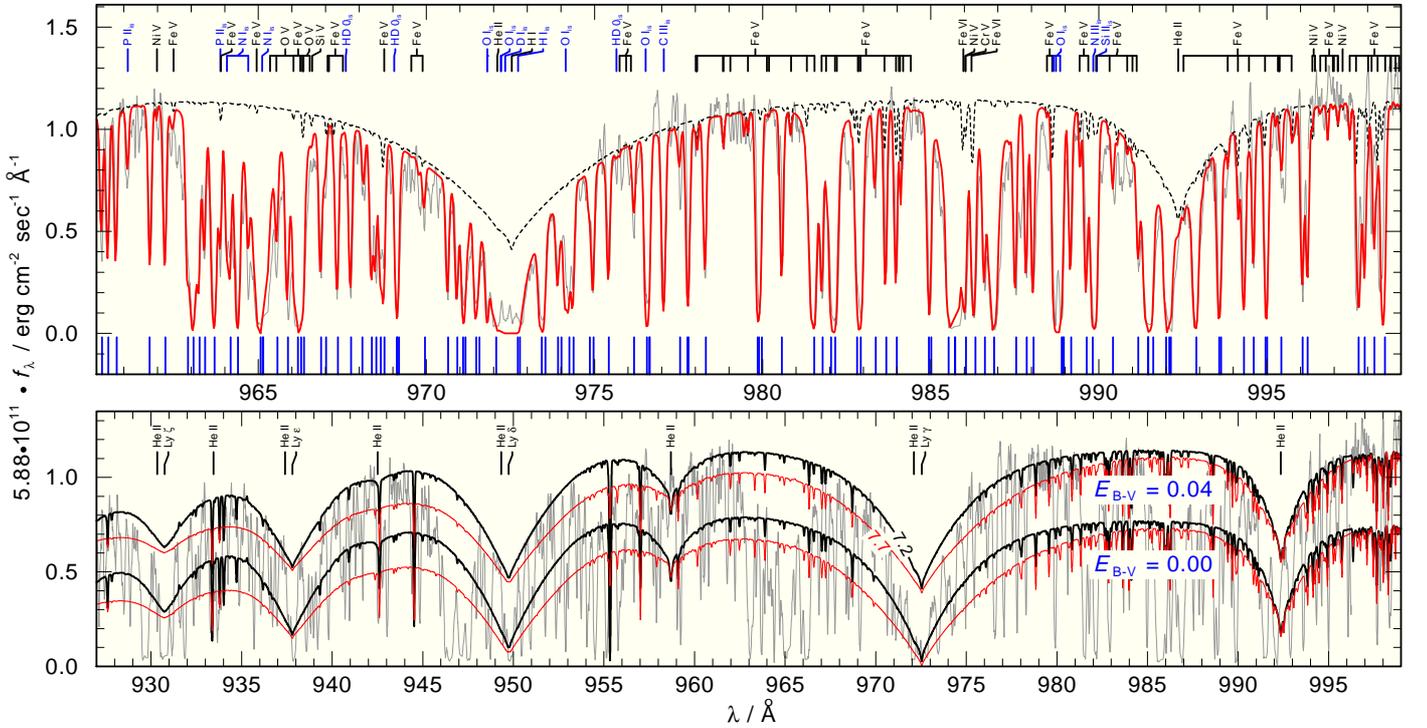}} 
    \caption{Top: Section of the \emph{FUSE} observation showing the necessity of
             a detailed modeling of the ISM. The dashed line is the final, pure photospheric
             model. The full (red) line shows a combined stellar and interstellar model.
             Blue marks at the bottom denote interstellar $\mbox{H}_2$ absorption lines.
             Bottom: Comparison of (pure photospheric) \emph{TMAP} models with \loggw{7.2} (thick, black)
             and \loggw{7.7} (thin, red), normalized to the flux of the \loggw{7.2} model
             at 999\,\AA\ for comparison. The full lines are calculated with the
             determined reddening of $E^\mathrm{max}_\mathrm{B-V} = 0.04$ (Sect.~\ref{sect:ismabsred}), 
             the dashed lines (shifted in $\log f_\lambda$ by -0.4 for clarity)
             with $E^\mathrm{min}_\mathrm{B-V} = 0$.
            }
   \label{fig:ism}
\end{figure*}

In the \emph{FUSE} wavelength range (Fig.~\ref{fig:ism}), 
numerous interstellar \mbox{H$_2$} and \Ion{H}{1} absorption lines hamper the analysis of
the photospheric spectrum. 
We applied the program \emph{OWENS}
to calculate ISM line-absorption models. 
\emph{OWENS} allows one to model different 
clouds with distinct radial and turbulent velocities, temperatures, column
densities, and chemical compositions. It fits Voigt profiles 
using a $\chi^2$ minimization. For more information on \emph{OWENS},
see e\@.g\@. \citet{hebrard02} or \citet{hebrard03}.    

Figure~\ref{fig:ism} shows a small
section of the \emph{FUSE} spectrum of \bd. 
Only the combination of the photospheric and ISM-model spectra 
reproduces most of its spectral features. 
Our ISM model includes lines of 
\Ion{H}{1}, 
H$_2$ ($J=0-9$), 
HD, 
\Ion{C}{2}\,--\Ion{}{4}, 
\Ion{N}{1}\,--\Ion{}{3}, \Ion{N}{5},
\Ion{O}{1}, \Ion{O}{6}, 
\Ion{Mg}{2}, 
\Ion{Al}{2},
\Ion{Si}{2}\,--\Ion{}{4}, 
\Ion{P}{2}, 
\Ion{S}{2}\,--\Ion{}{3}, \Ion{S}{6},
\Ion{Ar}{1}\,--\Ion{}{2}, 
\Ion{Mn}{2}, 
\Ion{Fe}{2}, and 
\Ion{Ni}{2}.
The radial velocity of the interstellar low-ionization ({\sc i} - {\sc ii})
atomic gas and H$_2$ is found to be $v_\mathrm{rad} = 6\pm 2\,\mathrm{km/s}$.

\section{Analysis}
\label{sect:ana}

In this section, we begin with a verification of those atmospheric
parameters that were already determined by \citet{herald02} which may
deviate because these authors used \emph{TLUSTY} and considered He, C, N, O, Si, and Fe,
while we employed \emph{TMAP} with far more elements (Tab\@.~\ref{tab:statistics}). 
This might result in a different atmospheric structure.
We begin with models with parameters of \citet{herald02} and fine-tune them
if necessary.

\subsection{Effective temperature and surface gravity}
\label{sect:ana:tefflogg}

\begin{figure}[ht!] 
   \resizebox{\hsize}{!}{\includegraphics{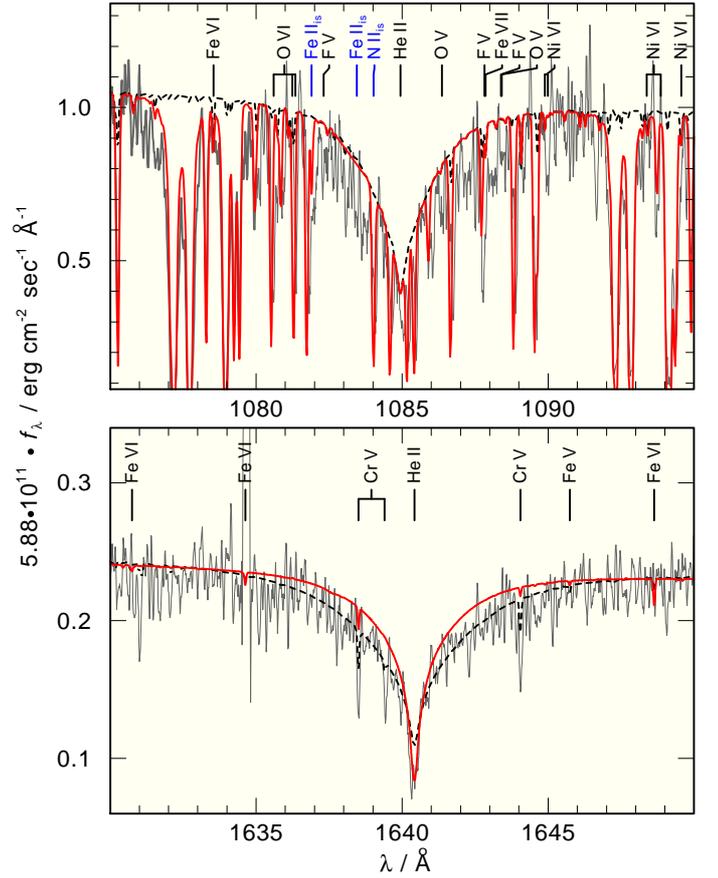}} 
    \caption{Spectra of our final model compared with the \emph{FUSE} and
             \emph{STIS} observations around
             \Ionw{He}{2}{1084.94} (top, dashed: pure photospheric,
                                         full line: photospheric + ISM line absorption)
             and 
             \Ionw{He}{2}{1640.42} (bottom, dashed: \loggw{7.7}, full line: \loggw{7.2}).
             The identified lines are marked. Most of the very strong absorptions in the
             top panel stem from H$_2$.
            }
   \label{fig:he}
\end{figure}

\citet{herald02} determined \Teffw{80} and \loggw{7.7}.
\emph{TMAP} test calculations showed that their \Teff\ value matches all ionization
equilibria well, e.g\@. 
\Ion{N}{4}\,/\,\Ion{N}{5},
\Ion{O}{4}\,/\,\Ion{O}{5}, and 
\Ion{Mn}{5}\,/\,\Ion{Mn}{6} (Sect\@.~\ref{sect:ana:metal}).
A lower \loggw{7.2} yields a significantly better fit agree of the theoretical \Ionw{He}{2}{1640.42} 
line profile with the observation in the line core, which is not matched at \loggw{7.7} (Fig.~\ref{fig:he}).
For the line-profile calculations, we used the Stark line-broadening tables by
\citet{schoeningbutler1989b,schoeningbutler1989a}.
The line wings are too broad at \loggw{7.7} and somewhat too weak at \loggw{7.2}.
The same is true for \Ionw{He}{2}{1084.94} (Fig.~\ref{fig:he}).
The decrements of spectral line series such as the 
\Ion{H}{1} Balmer \citep{rauchea98},
\Ion{H}{1} Lyman (transitions $n\,-\,n^\prime$ = 1\,--\,3 to 1\,--\,8), and
\Ion{He}{2} 2\,--\,$n^\prime$ series (2\,--\,5 to 2\,--\,11) 
are strongly dependent on \logg.
The fit to the outer line wings of the latter two series in the \emph{FUSE} spectrum
is also very well reproduced at \loggw{7.2}. 
Figure~\ref{fig:ism} shows
\Ion{He}{2} 2\,--\,5 to 2\,--\,14 and the respective
\Ion{H}{1} 1\,--\,3 to 1\,--\,7 blends.
A comparison of their theoretical profiles
at \loggw{7.2} and \loggw{7.7} with the observation
shows that they are significantly too broad at \loggw{7.7}. 
This effect cannot be compensated for by a lower $E_\mathrm{B-V}$ (Fig.~\ref{fig:ism}).
The interstellar \Ion{H}{1} absorption is not dominating the wings of these lines.
The observed ``shoulders'' between the \Ion{H}{1} Lyman lines and, thus,
the Lyman decrement is well reproduced at \loggw{7.2} and is 
definitely missed at \loggw{7.7}. This is best visble between
Ly\,$\zeta$ and Ly\,$\delta$ because the contamination of the
\emph{FUSE} observation with ISM lines is obviously weak. 
\loggw{7.2} agrees with the value of \citet{herald02} within \emph{realistic}
error limits of 0.3\,dex. Their assumed statistical error (\loggw{7.7^{+0.13}_{-0.18}}) seems to 
be too optimistic. We finally adopted \Teffw{80\pm 10} and \loggw{7.2\pm 0.3} for our analysis.

\subsection{Hydrogen and helium}
\label{sect:ana:hhe}

We adopted $\mathrm{[He]}=-0.3$ 
($[\mathrm{X}]$ denotes log [mass fraction / solar mass fraction] for species X) 
from \citet{herald02} and achieved a reasonable fit to the \Ion{He}{2} lines \sA{fig:he}.

\subsection{Metal abundances}
\label{sect:ana:metal}

In our \emph{TMAP} model-atmosphere calculations,
we considered opacities of all elements from H to Ni except for Li, Be, B, Cl, and K.
We convolved our synthetic spectra with Gaussians to simulate 
the respective instrument resolution.
By comparison with the observed spectrum of \bd,
we determined for the first time photospheric abundances of Ar, Cr, Mn, Co, and Ni. 
All identified lines of these species in the available observations
(Sect.~\ref{sect:obs}) were evaluated in our abundance analysis.
No lines of  F, Ne, Na, Mg, Al, Ca, Sc, Ti, and V were identified. 
For these, only upper limits could be derived by test models where the respective
lines in the model emerge (at the abundance limit) from the noise in 
the observation. 
A detailed comparison of models with different abundances of individual metals showed
that the typical abundance error is $\pm 0.3$\,dex.
In the following, we briefly describe some measurements (ordered by increasing atomic weight).

\begin{figure}[ht!] 
   \resizebox{\hsize}{!}{\includegraphics{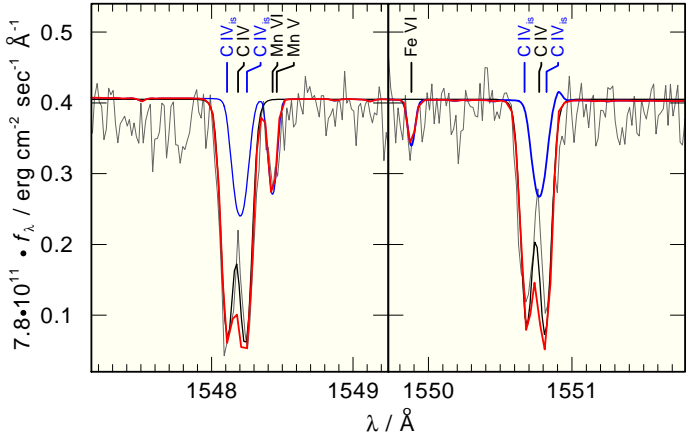}} 
    \caption{\Ionww{C}{4}{1548.20,1550.77} lines in the \emph{STIS}
             observation compared with our final model.
             Blue,  thin: photospheric                lines, 
             red,  thick: photospheric + interstellar lines,
             black, thin:                interstellar lines.
            }
   \label{fig:c}
\end{figure}

\paragraph{Carbon} lines of stellar origin are not detectable in the
spectrum of \bd.  Narrow interstellar \Ionww{C}{4}{1548.20,1550.77} lines
are present (Fig.~\ref{fig:c}) that virtually ``bracket'' the photospheric lines 
($v^\mathrm{star}_\mathrm{rad} = -14.4\,\mathrm{km/s}$, \se{sect:radvel}).
They have their origin in 
two different interstellar ``clouds'' with 
different radial velocities and column densities
(cloud~1: $v_\mathrm{rad} = -5.4\,\mathrm{km/s}$,
    $\log (n_\mathrm{C\,IV} / \mathrm{(g/cm^2)}) = 16.8$;
 cloud~2: $v_\mathrm{rad} = -30.4\,\mathrm{km/s}$,
    $\log (n_\mathrm{C\,IV} / \mathrm{(g/cm^2)}) = 14.6$). 
The radial velocity of cloud~1 agrees with the velocity of
the low-ionization ISM gas (Sect\@.~\ref{subsect:mod:ism}).
Interestingly, \citet{schneideretal83} determined 
$v_\mathrm{rad} = -6.6\pm 3.8\,\mathrm{km/sec}$ for the nebula.
This even suggests a nebular origin of cloud~1
while cloud~2 with its significantly higher velocity lies on the line
of sight toward us.

It is worthwhile to note that no other interstellar two-component absorption-line
feature is identified in the spectrum of \bd.

However, a photospheric component, even at low abundance, worsens the agreement
between model and observation in the shoulder between the interstellar
lines (Fig.~\ref{fig:c}).
Since the strongest \Ion{C}{4} lines in the \emph{FUSE} wavelength
range ($\lambda\lambda 948, 1107, 1168$\,\AA) become too strong
for $\mathrm{[C]} > -2.9$, we determine this as an upper limit. 
This value is about four times lower than that found by \citet{herald02}.

\paragraph{Nitrogen} \Ionw{N}{4}{1718.55} and \Ionww{N}{5}{1238.82, 1242.80} are prominent in the
\emph{STIS} observation. Our final model reproduces the strengths of these lines and
the ionization equilibrium at $\mathrm{[N]}= -1.7$ \sA{fig:n} well.

\begin{figure}[ht!] 
   \resizebox{\hsize}{!}{\includegraphics{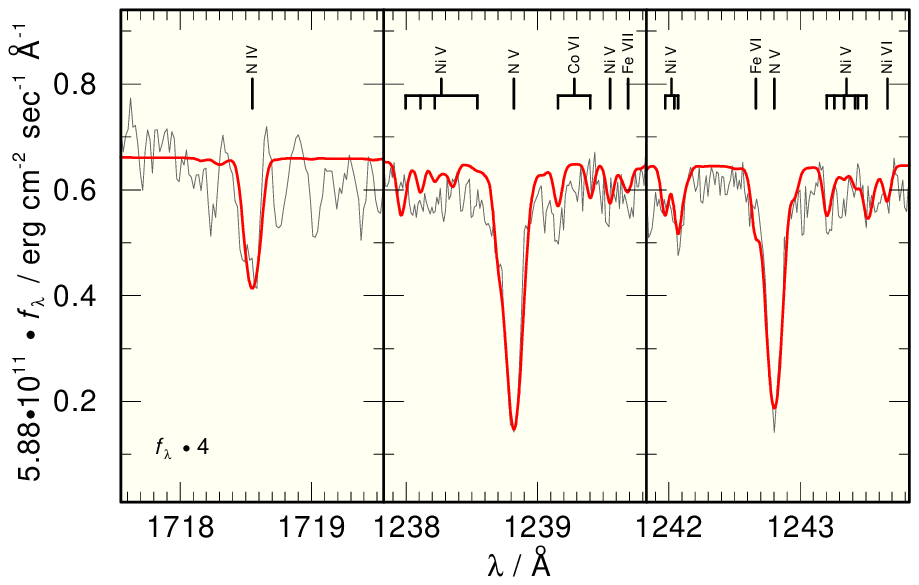}} 
    \caption{\Ionw{N}{4}{1718.55} and \Ionww{N}{5}{1238.82, 1242.80} lines in the \emph{STIS} 
             observation compared with our final model.
            }
   \label{fig:n}
\end{figure}

\citet{herald02} determined a much lower photospheric N abundance 
($< 10^{-3}$ times the solar value)
because of strong line wings forming at higher abundances.
The reason may be a different approximation by the quadratic Stark effect 
in \emph{TMAP} and \emph{TLUSTY}/\emph{SYNSPEC}.

For the line-absorption cross-sections \emph{TMAP} calculates

\begin {equation}
\sigma_\mathrm{ij}(\nu) = 
               \frac{\sqrt{\pi}e^2}{m_\mathrm{e}c} 
               \frac{f_\mathrm{ij} }{\Delta \nu _\mathrm{D}} ,
               H(a,\varv)
\end{equation}

\noindent
$e$ and $m_\mathrm{e}$ are electron charge and mass,
$c$ is the light velocity, 
$\Delta\nu_\mathrm{D}$ the Doppler width, 
$f_\mathrm{ij}$ the oscillator strength for the transition i $\rightarrow$ j,
and
$H$ the Voigt function
with

\begin {equation}
a= \frac{\Gamma}{4\pi\Delta \nu _\mathrm{D}} ,
\end{equation}

\noindent
and
\begin {equation}
\varv = \frac{\Delta \nu}{\Delta \nu _\mathrm{D}} ,
\end{equation}

\noindent
where $\Delta \nu = \nu - \nu_0$ with the line frequency $\nu_0$.

\noindent
According to \citet{cowley1970,cowley1971}, the line-broadening due to the 
quadratic Stark effect is approximately given by

\begin{equation}
\Gamma_{Stark} = 5.5\times10^{-5}\;\frac{n_e}{\sqrt{T}}\;\left
    [\,\frac{(n_\mathrm{eff}^\mathrm{up})^2}{z+1}\,\right]^2 ,
\end{equation}
where 
$n_\mathrm{e}$ is the electron density,
$T$ is the temperature,
$n_\mathrm{eff}^\mathrm{up}$ is the effective principal quantum number of the upper level, 
and 
$z$ is the effective charge seen by the active electron.
Radiative ($\Gamma_\mathrm{rad}$) and collisional damping by electrons is considered so that

\begin {equation}
\Gamma = \Gamma_\mathrm{rad} + \Gamma_\mathrm{Stark} .
\end{equation}

\noindent
\emph{TMAP} uses
$f_\mathrm{ij}$ of 
$7.8580\cdot 10^{-2}$ and 
$1.5716\cdot 10^{-1}$, 
and classical damping constants
$\Gamma_\mathrm{rad}$ of 
$1.4401\cdot 10^{+9}$ and 
$1.4494\cdot 10^{+9}$ for
\Ionww{N}{5}{1242.80, 1238.82}, respectively.

A comparison of calculated $\Gamma _\mathrm{St}$ values
(Sahal-Br\'echot priv\@. comm.)
has shown that 
$\Gamma_\mathrm{Stark}^\mathrm{Cowley\,}/\,\Gamma_\mathrm{Stark}^\mathrm{Sahal-Br\mathaccent"70B4{e}chot} = 0.33$
is a good approximation for the \Ion{N}{5} resonance doublet within $10\,000 < T/\mathrm{K} < 200\,000$.
We introduced this factor and adopted 
$\Gamma _\mathrm{St} = 3.0 \cdot \Gamma_\mathrm{Stark}^\mathrm{Cowley}$ 
for our analysis.

\begin{figure}[ht!] 
   \resizebox{\hsize}{!}{\includegraphics{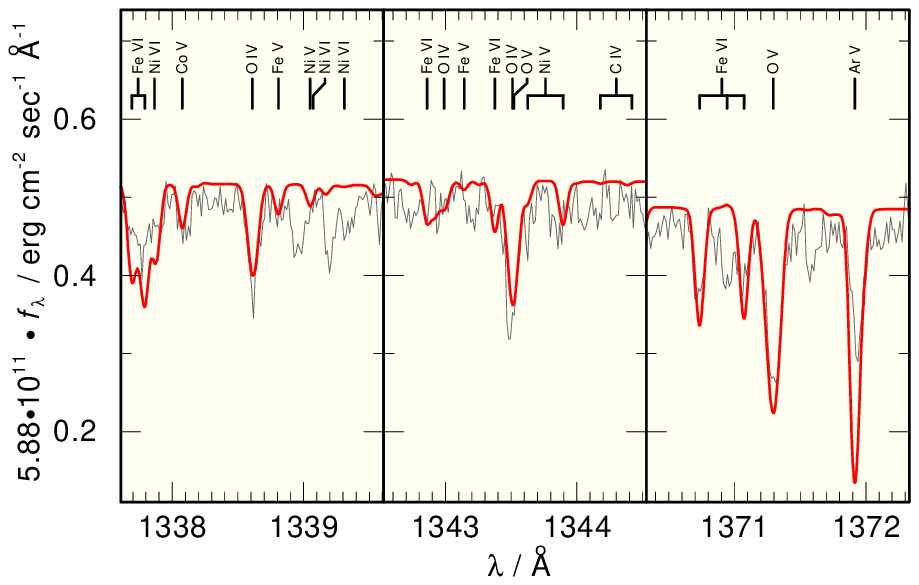}} 
    \caption{\Ionww{O}{4}{1338.61, 1343.51} (left and middle) and
             \Ionw{O}{5}{1371.29} (right) lines in the \emph{STIS} 
             observation compared with our final model.
            }
   \label{fig:o}
\end{figure}

\paragraph{Oxygen} is present in the spectra with the ionization stages 
{\sc iv} and {\sc v}.
We can reproduce the ionization equilibrium and the line strengths.
Figure~\ref{fig:o} shows 
\Ionww{O}{4}{1338.61, 1343.51} 
and 
\Ionw{O}{5}{1371.29} 
at $\mathrm{[O]}= -2.6$.
This agrees well with the observation 
and the value of \citet{herald02}.

\paragraph{Fluorine} was not identified in the observation. 
We modeled \Ionww{F}{5}{1082.31,1088.38} and \Ionw{F}{6}{1139.49}.
They would be visible in the observation at abundances
higher than solar.

\paragraph{Magnesium} lines are weak and almost fade in the noise,
even the strongest, \Ionw{Mg}{4}{1683.003}. 
Figure~\ref{fig:stis} shows \Ionw{Mg}{4}{1490.433}, which is not detectable in
the \emph{FUSE} observation at solar abundance.
Thus, we can only determine an upper
abundance limit of $\mbox{[Mg]} = 0.0$.

\begin{figure}[ht!] 
   \resizebox{\hsize}{!}{\includegraphics{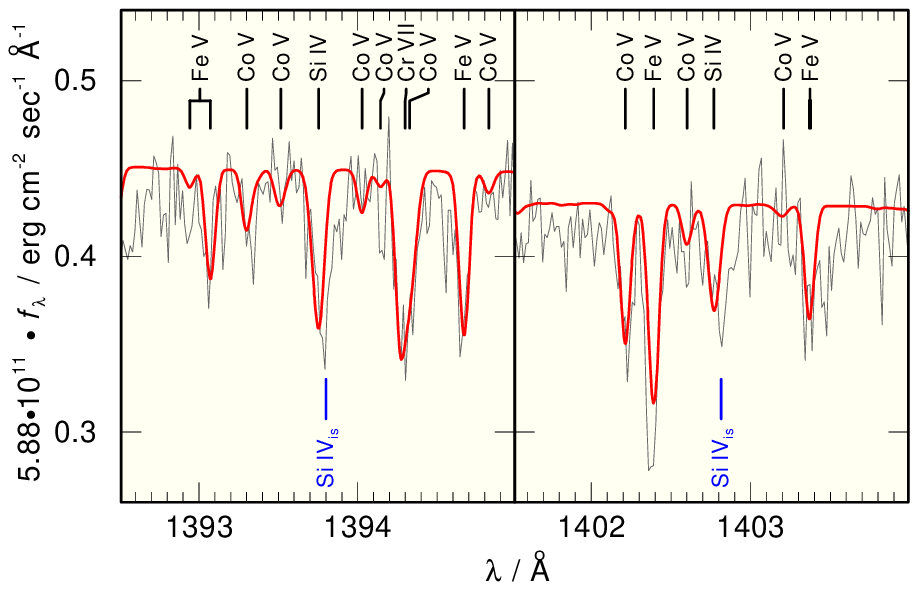}} 
    \caption{\Ionww{Si}{4}{1393.76,1402.77} in the \emph{STIS} 
             observation compared with our final model.
            }
   \label{fig:si}
\end{figure}

\paragraph{Silicon} exhibits only the \Ionww{Si}{4}{1393.76,1402.77} resonance doublet
(Fig.~\ref{fig:si}), other Si lines are too weak. A weak ISM contamination 
is present but we can determine $\mbox{[Si]} = -2.1$.
This agrees well with the upper limit of \citet{herald02}.

\begin{figure}[ht!] 
   \resizebox{\hsize}{!}{\includegraphics{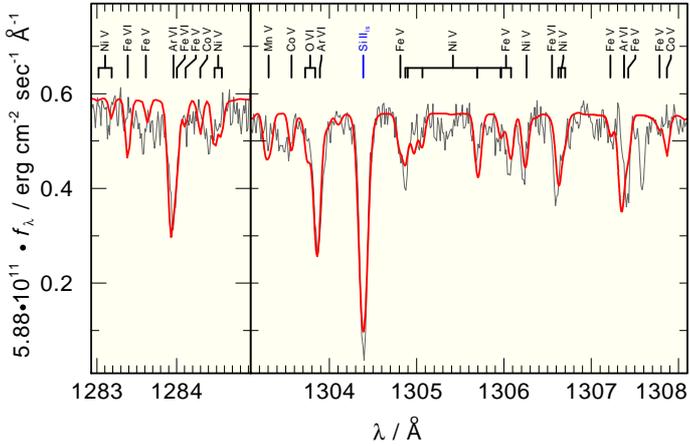}} 
    \caption{Section of the \emph{STIS} spectrum around 
             \Ionww{Ar}{6}{1283.96,1303.89,1307.38} compared with our
             final, composite photospheric and ISM spectrum.    
            }
   \label{fig:ar}
\end{figure}

\paragraph{Argon} is strongly enhanced ($\mathrm{[Ar]} = +1.1$).
Similar to the central star of the PN \object{Sh\,2--216},
\object{LS\,V $+46^\circ 21$} \citep[cf\@.][]{rauch07}, 
\Ionww{Ar}{6}{1283.96, 1303.89, 1307.38} are identified in the \emph{STIS} observation 
(Fig\@. \ref{fig:ar}) and reproduced by our final model. 
Our models show a strong \Ionw{Ar}{7}{1063.55} line but this
is blended by interstellar $\mbox{H}_2$ absorption and therefore not suitable
for a \Teff\ determination via the ionization equilibrium.

A problem are lines like \Ionw{Ar}{5}{1371.92} (Fig.~\ref{fig:o}) which 
are much stronger in our models than observed. Since many ionization equilibria of 
different species such as Mn {\sc v} - {\sc vi} (see below) are well-matched, a higher 
\Teff\ that would shift the ionization balance from \Ion{Ar}{5} to \Ion{Ar}{6} can
be excluded. 
We checked the possibility that the oscillator strengths of \Ion{Ar}{5}
lines that we use in \emph{TMAP} \citep[from the Opacity Project,][]{seaton94} are too strong.
The \Ion{Ar}{5} triplet 3p$^3$ $^3$S$^o$ -- 3p$^4$ $^3$P 
(\Ionww{Ar}{5}{1341.57, 1350.39, 1371.92}) is found in OpacityProject data
(far off) at 1455.77\,\AA\ (without fine-structure splitting) with 
$f = 4.2333\cdot 10^{-2}$.
From this value, we calculated $f$-values of 
$4.7037\cdot 10^{-2}$, 
$1.4111\cdot 10^{-2}$, and 
$2.3518\cdot 10^{-2}$
for the three fine-structure components.
A comparison with recent work of 
\citet[][$f_{1371.92} = 2.257\cdot 10^{-2}$ and $2.834\cdot 10^{-2}$]{tayaletal09} 
shows good agreement with the OpacityProject. Thus, the reason for the disagreement of
the \Ion{Ar}{5} lines remains unknown.

\begin{figure}[ht!] 
   \resizebox{\hsize}{!}{\includegraphics{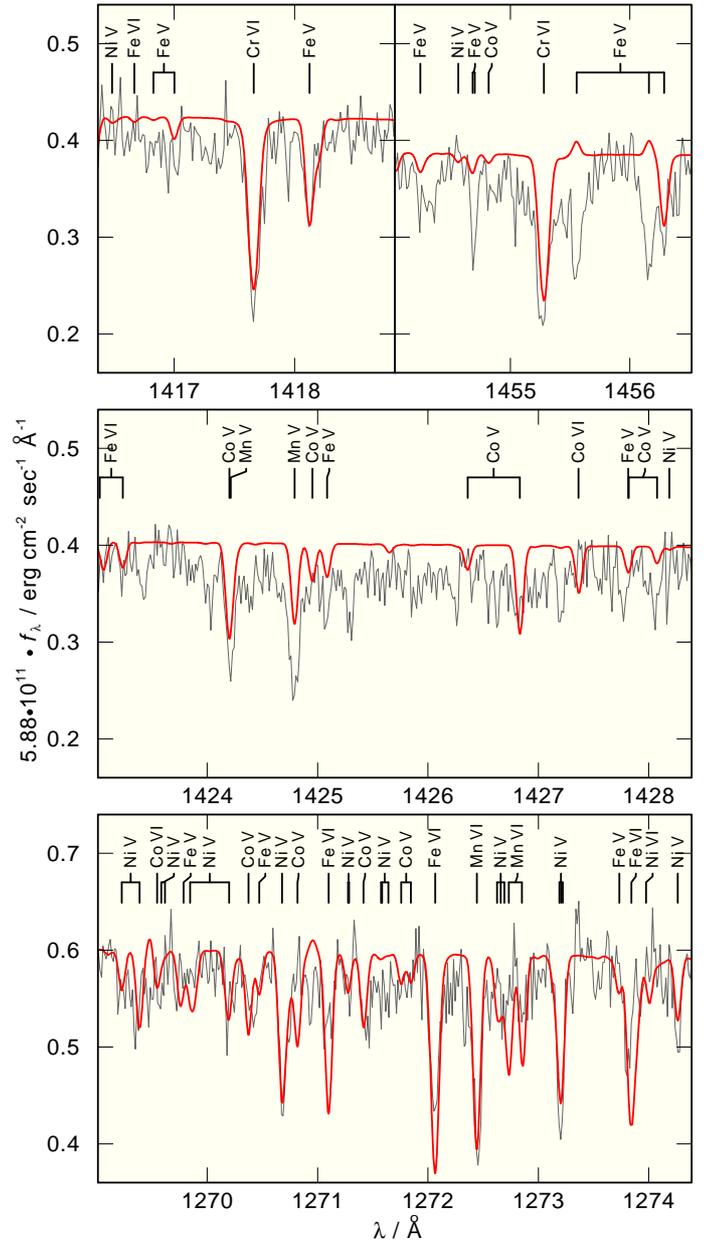}} 
    \caption{Sections of the \emph{STIS} observation containing (in addition to others) lines
      of \Ion{Cr}{6} (top), Co (middle), and Ni (bottom) compared with our final model.      
            }
   \label{fig:crconi}
\end{figure}

\paragraph{Chromium} is about 70 times enhanced ($\mbox{[Cr]} = +1.8$).
In Fig.~\ref{fig:crconi}), \Ionww{Cr}{6}{1417.66, 1455.28} are shown that are
well-reproduced by our final model.  
\Ionww{Cr}{5}{1482.76, 1489.71} (Fig.~\ref{fig:stis}) would require a much higher
$\mbox{[Cr]} = +3.1$ (Sect.~\ref{subsec:photo}). Since \Teff\ and \logg\ are well-determined
(Sect.~\ref{sect:ana:tefflogg}) and, thus, a strong change of the ionization balance
toward \Ion{Cr}{5} is impossible within the error limits, we have to consider these
two lines as blends and not suited for a reliable abundance determination.

\begin{figure*}[ht!] 
   \resizebox{\hsize}{!}{\includegraphics{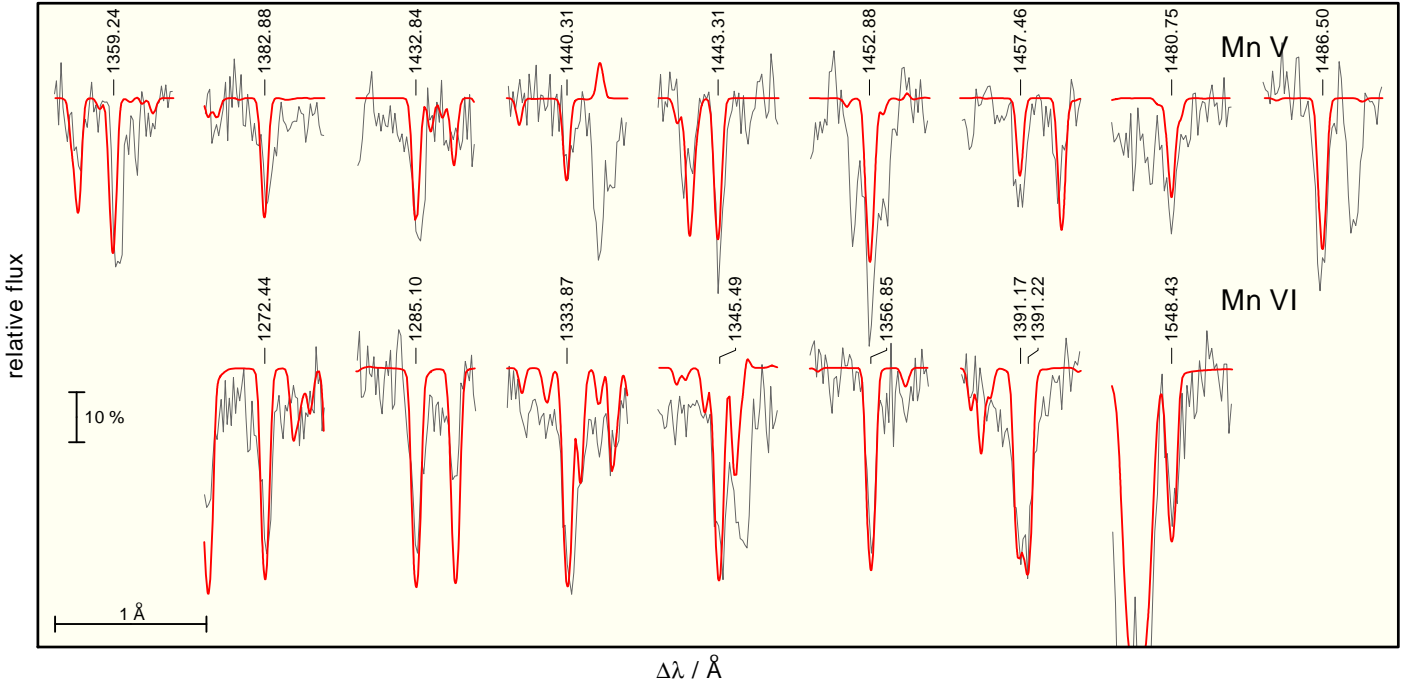}} 
    \caption{Strongest \Ion{Mn}{5} (top) and \Ion{Mn}{6} lines (bottom) in the \emph{STIS}
             wavelength range compared with our final model.      
            }
   \label{fig:mn}
\end{figure*}

\paragraph{Manganese} exhibits several \Ion{Mn}{5} and \Ion{}{6} lines in the \emph{STIS} 
wavelength range, e\@.g\@. (the strongest)
\Ion{Mn}{5} $\lambda\lambda$ 1359.24, 1382.88, 1432.84, 1440.31, 1443.31, 
              1452.88, 1457.46, 1480.75, 1486.50\,\AA\ and 
\Ion{Mn}{6} $\lambda\lambda$ 1272.44, 1285.10, 1333.87, 1345.49, 1356.85, 1391.17, 1391.22, 1548.43\,\AA\
(Fig.~\ref{fig:mn}). 
All these lines are reproduced at $\mbox{[Mn]} = +1.5$.
Since many lines of two successive ionization stages are suitable for
spectral analysis, the Mn {\sc v} - {\sc vi} ionization equilibrium is one of the
prime indicators for \Teff\ for \bd. \Teffw{80} is well-matched at \loggw{7.2}
(Sect.~\ref{sect:ana:tefflogg}).

\paragraph{Iron} has many lines of \Ion{Fe}{5} and \Ion{}{6} in the UV spectrum.
In contrast to \citet{herald02}, who determined a $0.1 - 0.5$ times solar Fe abundance, 
we find it approximately solar (e.g\@. Fig.~\ref{fig:stis}). There are, however,
some Fe lines that are stronger in our model than observed (see some \Ion{Fe}{6} lines
in \ab{fig:crconi}, bottom panel). Within the error range,
we find the same Fe abundance as \citet{herald02}.

\paragraph{Cobalt} is strongly enhanced.
\Ionww{Co}{5}{1331.50,1349.60} 
indicate $\mbox{[Co]} = +2.2$ \sA{fig:crconi}.

\paragraph{Nickel} is also overabundant. Numerous \Ion{Ni}{5} and \Ion{}{6} lines are found in the UV
spectrum. They are all reproduced at $\mathrm{[Ni]} = +0.7$ (e.g\@. Figs.~\ref{fig:stis}, \ref{fig:crconi}).

\subsection{Mass, luminosity, radius, and distance}
\label{sect:mass}

\begin{figure}[ht!] 
   \resizebox{\hsize}{!}{\includegraphics{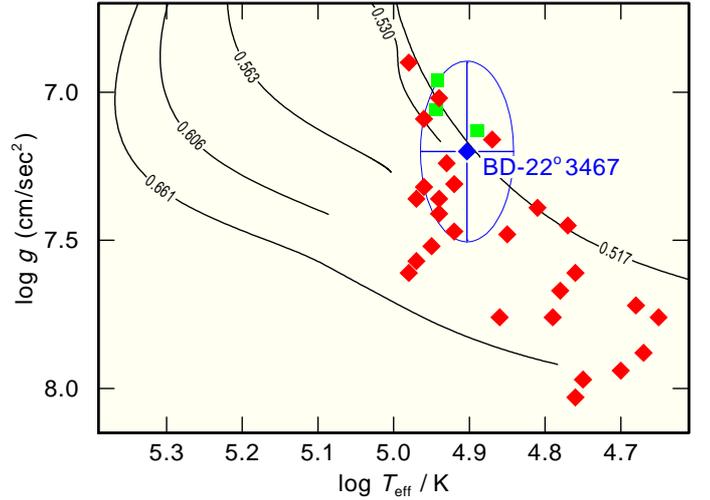}} 
    \caption{Location of \bd\ in the $\log$
             \Teff\,--\,\logg\ diagram compared with other DAO-type WDs taken from 
             \citet[all symbols]{gianninas10}.
             Three of these (green squares) are used to compare their
             metal abundances with those of \bd\ (Fig.~\ref{fig:x}).
             The post-AGB evolutionary tracks were calculated by Miller Bertolami 
             (priv\@. comm.) labeled with the stellar mass (in $M_\odot$). 
            }
   \label{fig:evolpagb}
\end{figure}

We determined the stellar mass and luminosity from a comparison with
evolutionary tracks 
for hydrogen-rich post-AGB stars (Fig.~\ref{fig:evolpagb}, Miller Bertolami, priv\@. comm.)
and
for post-EHB (extended horizontal branch)
stars from \citet[Fig.~\ref{fig:evolpehb}]{dormanetal93}. 
The results are summarized in Tab.~\ref{tab:mlrd}.

The stellar radius follows from $R = \sqrt{G\,M\,/\,g}$, where $G$ is the gravitational constant.
We determined the spectroscopic distance $D = 361^{+195}_{-137}\,\mathrm{pc}$ (Table~\ref{tab:mlrd})
following \citet{wassermann10}.
It is derived from
$$ D/\mathrm{pc} = 6.6175 \cdot 10^{-6} \sqrt{F_\nu/F_\mathrm{obs} \cdot M/M_{\odot} \cdot 10^{-\log g}}$$ 
with the astrophysical flux $F_\nu$ at the surface of the star as predicted by the model spectrum, 
and $F_\mathrm{obs}$ is the observed, extinction-corrected flux
(we used $F_\mathrm{1513\,\AA}$ to match the center of the \emph{GALEX} FUV band).
$D$ is about three times larger than the improved \emph{HIPPARCOS} distance
($D = 119^{+28}_{-19}\,\mathrm{pc}$, Sect.~\ref {sect:intro}) but it agrees with
the photometric distance \citep[$360\pm 80 \,\mathrm{pc}$,][]{jacoby81}
for the cool component.

\begin{table}[ht!]\centering
\caption{Mass, luminosity, radius, distance, and height above the Galactic plane of \bd.}         
\label{tab:mlrd}
\begin{tabular}{rcc}
\hline
\hline
                               & post-AGB evolution        & post-EHB evolution        \\
\hline
\noalign{\smallskip}
$M\,/\,M_\odot$       & $0.52\pm 0.05$            & $0.48\pm 0.05$            \\ 
$\log\ L\,/\,\mathrm{L_\odot}$ & $1.50^{+0.35}_{-0.45}$    & $1.00^{+0.50}_{-0.80}$    \\
\noalign{\smallskip}
$R\,/\,\mathrm{R_\odot}$       & $0.030^{+0.013}_{-0.009}$ & $0.029^{+0.012}_{-0.009}$ \\
\noalign{\smallskip}
$D\,/\,\mathrm{pc}$            & $376^{+203}_{-142}$       & $361^{+195}_{-137}$       \\
\noalign{\smallskip}
$H\,/\,\mathrm{pc}$            & $233^{+53}_{-66}$         & $224^{+51}_{-62}$         \\
\noalign{\smallskip}
\hline
\end{tabular}
\end{table}

It is worthwhile to mention that there is no agreement with the \emph{HIPPARCOS} distance
at the maximum value of \loggw{7.83} given by \citet{herald02} either. 
The calculated spectroscopic distance 
(using $M = 0.68\,M_\odot$, $R = 0.017\,\mathrm{R_\odot}$, and the respective model flux)
is then $D = 226\,\mathrm{pc}$, about twice the \emph{HIPPARCOS} distance.
Although trigonometric parallaxes in general have to be regarded as hard constraints,
orbital movement in binary systems such as \bd\ may bear on the precision of 
\emph{HIPPARCOS} measurements \citep[cf\@.][for \object{LB\,3459} = \object{TYC\,9166$-$00716$-$1}]{rauch00}.
Figures \ref{fig:ism} and \ref{fig:he} clearly demonstrate that the spectra cannot be reproduced with a 
gravity of \loggw{7.7} or higher, while a good fit is achieved for \loggw{7.2}.
The upcoming \emph{GAIA} mission will provide a highly precise parallax for
\bd\ which will be able to rule out the uncertainties.

\section{Results and discussion}
\label{sect:discussion}

We calculated a grid of \emph{TMAP} NLTE models atmospheres 
for \bd, the exciting star of the ionized nebula \object{A\,35}. 
We determined \Teffw{80\pm 10} and \loggw{7.2 \pm 0.3}, and thus confirm
\Teffw{80} found by \citet{herald02}. Nevertheless, we found a
somewhat lower \logg\ than \citet[\loggw{7.7}]{herald02}.
For the first time in this object, abundances from detailed line-profile fits 
were determined for Ar, Cr, Mn, Co, and Ni (Tab.~\ref {tab:results}). 
For F, Ne, Na, Mg, Al, Ca, Sc, Ti, and V, upper limits could be derived.

\begin{table}[ht!]\centering
\caption{Results of this work compared with \citet{herald02},
         whose He abundance we adopted.}         
\label{tab:results}
\renewcommand{\tabcolsep}{1mm}
\begin{tabular}{llcc}
\hline
\hline
\noalign{\smallskip}
&                                                                   & this work                   & H \& B                       \\
\hline         
\noalign{\smallskip}
& NLTE code                                                         &  \emph{TMAP}                & \emph{TLUSTY}                \\  
\hline         
\noalign{\smallskip}
\it{primary} 
&$T_\mathrm{eff}\,/\,$kK                                            &  $80\pm 10$                 & $80\pm 3$                    \\
&$\log\ g$                                                          &  $7.2\pm 0.3$               & $7.7^{+0.13}_{-0.18}$        \\
\noalign{\smallskip}
&\multicolumn{1}{r}{$\mathrm{X_{He}(X_\odot)}$}\hspace{6mm}\hbox{}  &  $0.47$                     & $0.3-0.5$                    \\
&\multicolumn{1}{r}{$\mathrm{X_{C}(X_\odot)}$ }\hspace{6mm}\hbox{}  &  $1.2\times 10^{-3}$        & $5.0\times 10^{-3}-10^{-2}$  \\
&\multicolumn{1}{r}{$\mathrm{X_{N}(X_\odot)}$ }\hspace{6mm}\hbox{}  &  $1.8\times 10^{-2}$        & $< 10^{-3}$                  \\
&\multicolumn{1}{r}{$\mathrm{X_{O}(X_\odot)}$ }\hspace{6mm}\hbox{}  &  $2.5\times 10^{-3}$        & $5.0\times 10^{-3}-10^{-2}$  \\
&\multicolumn{1}{r}{$\mathrm{X_{F}(X_\odot)}$ }\hspace{6mm}\hbox{}  &  $\sla 1$                   &                              \\
&\multicolumn{1}{r}{$\mathrm{X_{Ne}(X_\odot)}$}\hspace{6mm}\hbox{}  &  $< 1$                      &                              \\
&\multicolumn{1}{r}{$\mathrm{X_{Na}(X_\odot)}$}\hspace{6mm}\hbox{}  &  $< 1$                      &                              \\
&\multicolumn{1}{r}{$\mathrm{X_{Mg}(X_\odot)}$}\hspace{6mm}\hbox{}  &  $< 1$                      &                              \\
&\multicolumn{1}{r}{$\mathrm{X_{Al}(X_\odot)}$}\hspace{6mm}\hbox{}  &  $< 1$                      &                              \\
&\multicolumn{1}{r}{$\mathrm{X_{Si}(X_\odot)}$}\hspace{6mm}\hbox{}  &  $7.5\times 10^{-3}$        & $< 10^{-2}$                  \\
&\multicolumn{1}{r}{$\mathrm{X_{P}(X_\odot)}$ }\hspace{6mm}\hbox{}  &  $< 1.4\times 10^{-3}$\hspace{3mm}\hbox{}     &                              \\
&\multicolumn{1}{r}{$\mathrm{X_{S}(X_\odot)}$ }\hspace{6mm}\hbox{}  &  $< 1.2\times 10^{-3}$\hspace{3mm}\hbox{}       &                              \\
&\multicolumn{1}{r}{$\mathrm{X_{Ar}(X_\odot)}$}\hspace{6mm}\hbox{}  &  $12$                       &                              \\
&\multicolumn{1}{r}{$\mathrm{X_{Ca}(X_\odot)}$}\hspace{6mm}\hbox{}  &  $< 1$                      &                              \\
&\multicolumn{1}{r}{$\mathrm{X_{Sc}(X_\odot)}$}\hspace{6mm}\hbox{}  &  $< 1$                      &                              \\
&\multicolumn{1}{r}{$\mathrm{X_{Ti}(X_\odot)}$}\hspace{6mm}\hbox{}  &  $< 1$                      &                              \\
&\multicolumn{1}{r}{$\mathrm{X_{V}(X_\odot)}$ }\hspace{6mm}\hbox{}  &  $< 1$                      &                              \\
&\multicolumn{1}{r}{$\mathrm{X_{Cr}(X_\odot)}$}\hspace{6mm}\hbox{}  &  $70$                       &                              \\
&\multicolumn{1}{r}{$\mathrm{X_{Mn}(X_\odot)}$}\hspace{6mm}\hbox{}  &  $35$                       &                              \\
&\multicolumn{1}{r}{$\mathrm{X_{Fe}(X_\odot)}$}\hspace{6mm}\hbox{}  &  $1$                        & $0.1-0.5$                    \\
&\multicolumn{1}{r}{$\mathrm{X_{Co}(X_\odot)}$}\hspace{6mm}\hbox{}  &  $150$                      &                              \\
&\multicolumn{1}{r}{$\mathrm{X_{Ni}(X_\odot)}$}\hspace{6mm}\hbox{}  &  $5$                        &                              \\
&\ebv                                                               &  $0.02\pm 0.02$             & $0.04\pm 0.01$               \\
&$\log \nh$                                                         &  $20.7\pm 0.1$\hspace{1.5mm}\hbox{}              & $20.9\pm 0.1$\hspace{1.5mm}\hbox{}                \\
&$D^\mathrm{a}\,/\,\mathrm{pc}$                                 &  $361^{+195}_{-137}$        & 163 (\emph{HIPPARCOS})       \\
&$M^\mathrm{a}\,/\,M_\odot$                            &  $0.48\pm 0.05$             & $0.5^{+0.5}_{-0.4}$          \\
&$R^\mathrm{a}\,/\,\mathrm{R_\odot}$                            &  $0.029^{+0.012}_{-0.009}$  & $0.0165^{+0.01}_{-0.006}$    \\
\noalign{\smallskip}
&$R^\mathrm{a}/D^\mathrm{a}\,/\,\mathrm{R_\odot\,/\,\mathrm{kpc}}$ &  $0.080^{+0.004}_{-0.001}$  & $0.1\pm 0.01$                \\
\noalign{\smallskip}                                                
&$\log\ L^\mathrm{a}\,/\,\mathrm{L_\odot}$                      &  $1.00^{+0.50}_{-0.80}$     & $1.0\pm 0.4$                 \\
\noalign{\smallskip}
\hline         
\noalign{\smallskip}
\it{secondary} 
&$T_\mathrm{eff}\,/\,$kK                                            &  $5$                        & $5$                          \\
&$\log\ g$                                                          &  $3.5$                      & $3.5$                        \\
\noalign{\smallskip}
\hline
\end{tabular}
\begin{list}{}{}
\item[$^{\mathrm{a}}$] our values for post-EHB evolution
\end{list}
\end{table}

\begin{figure}[ht!] 
   \resizebox{\hsize}{!}{\includegraphics{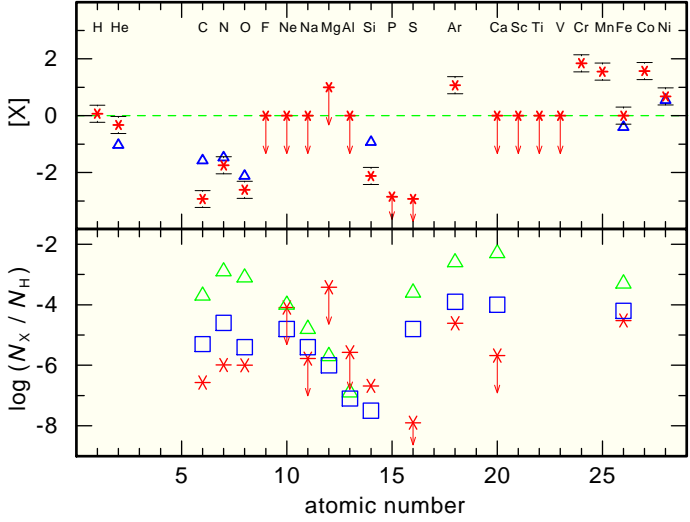}} 
    \caption{Top: Photospheric abundances of \bd\ compared with solar values
             \citep{asplund09}, and mean values from three objects of \citet[blue triangles]{good05} 
             that are located within the error ellipse (cf\@. Fig.~\ref{fig:evolpagb}, green squares there). 
             Arrows indicate upper limits.
             Bottom: Comparison of number ratios compared with predictions of
             diffusion calculations for 
             hydrogen-rich (DA-, blue squares) and 
             helium-rich (DO-type, green triangles) WDs \citep{cea1995, cfw1995} with \Teffw{80} and \loggw{7.2}.     
            }
   \label{fig:x}
\end{figure}

The element abundance pattern (Fig.~\ref{fig:x}) is probably the result of the 
interplay of radiative levitation and gravitational settling.
Compared to those objects in the DAO sample from \citet{good04, good05}
that have the same \Teff\ and \logg\ within error limits,
\bd\ has a higher He and a lower Si abundance.
Cr, Mn, Co, and Ni \citep[not measured by][]{good05} are strongly overabundant,
likely due to the force of the radiation field. 
Radiative levitation may be a reason but the question remains why Fe is about solar. 
Detailed diffusion calculations that consider these elements are highly desirable
investigating this phenomenon.
A comparison with existing diffusion calculations 
\citep[Fig.~\ref{fig:x}]{cea1995, cfw1995} shows
(within error limits) that the metal abundances partially agree with predictions for DA-type WDs.
Only N, Mg, S, and Ca are more than one dex lower than predicted.

Until recently, \bd\ was believed to be the central star of a planetary nebula
and, thus, a post-AGB evolution of an intermediate-mass star
($0.8\,M_\odot\,\sla\,M_\mathrm{initial}\,\sla\,8.0\,M_\odot$)
appeared to be a matter of course.
\citet{frew10} showed that the visible, PN-like nebula may be a combination of
shock excitation and photoionization of the ambient interstellar gas.

\bd\ provides enough high-energy photons, 
($4.4\cdot 10^{25}\,\mathrm{photons/s/cm^2}$ at energies $>$\,13.6\,eV and
 $6.4\cdot 10^{22}\,\mathrm{photons/s/cm^2}$ at energies $>$\,54\,eV
calculated from our final model which give totals of 
$N_{13} = 2.3\cdot 10^{45}\,\mathrm{photons/s}$ and
$N_{54} = 3.3\cdot 10^{42}\,\mathrm{photons/s}$, respectively)
to ionize the surrounding interstellar gas \citep[cf\@.][]{rauchea04}.
For $N_{54} / N_{13} < 0.01$, no \Ion{He}{2} emission from the nebula can be 
expected. If we assume an electron temperature of $T_\mathrm{e} = 10\,000\,\mathrm{K}$
in the nebula, negligible extinction, $d = 361\,\mathrm{pc}$ (Table~\ref{tab:results}),
and an average angular diameter of 750\,\arcsec, we derive that \bd\ has a 
Str\"omgren sphere with a density of 16\,cm$^{-3}$, 
$\log (F_\mathrm{H\,\beta} / \mathrm{erg/s/cm^2}) = -11.23$, and an ionized nebular mass of 
0.5\,M$_\odot$. The absolute H\,$\beta$ flux agrees very well with the value given by
\citet[][$\log F_\mathrm{H\,\beta} = -11.3$]{ackeretal1992}.

The nebular spectrum by Acker \& Stenholm obtained for the catalog \citep{ackeretal1992}
is weak and, consequently, the $F_\mathrm{H\,\alpha} / F_\mathrm{H\,\beta}$ ratio is overestimated,
and the derived extinction of $c = 0.7$ is very uncertain.
$F_{[O\,III] \lambda 5007\,\AA} / F_\mathrm{H\,\beta} = 2$ indicates low excitation.
Both \Ionww{N}{2}{6548, 6584} are visible and, with standard assumptions for the
nebula but without any correction for unseen higher ionization stages of nitrogen, its
nebular abundance may be estimated to be $12 + \log N/H  = 7.7$.
The \Ionw{Si}{2}{6716} / \Ionw{Si}{2}{6731} line intensity ratio is at the low-density limit
and we can conclude that the nebular electron density is $n_\mathrm{e} < 100\,\mathrm{cm^3}$.
In addition, we may estimate that the nebular oxygen and sulfur abundances are 7.9 and
6.4, respectively. 
It is worthwhile to note that the ionized mass of 0.5\,M$_\odot$ is of about
the value usually assumed for the average PN mass of 0.2 - 0.3\,M\,$_\odot$.
The reason is simply the fairly low luminosity of \bd. Such a low-luminosity ionizing star 
makes a Str\"omgren sphere that may well mimic a genuine PN. It is thus possible that similar 
objects have been misclassified as PNe.

\begin{figure}[ht!] 
   \resizebox{\hsize}{!}{\includegraphics{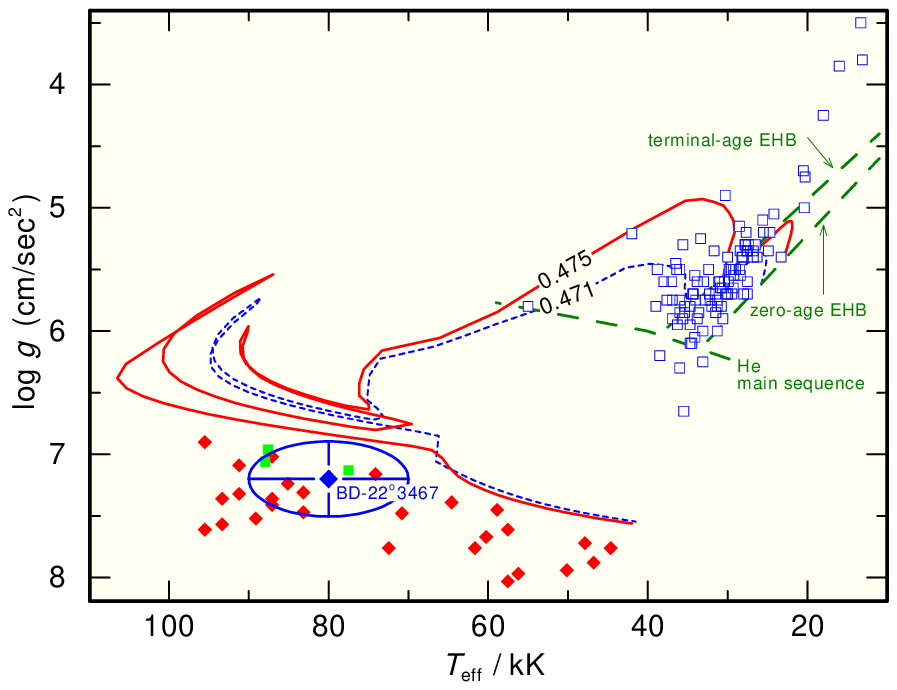}} 
    \caption{Location of \bd\ in the
             \Teff\,--\,\logg\ diagram compared with 
             sd(O)B-type stars close to the EHB taken from \citet[blue, open squares][]{edelmann03} and
             DAO-type WDs taken from  \citet[red diamonds and green, filled squares][cf\@. Fig.~\ref{fig:evolpagb}]{gianninas10}.
             The post-EHB evolutionary tracks are from \citet[$Y = 0.288 \approx Y_\odot$]{dormanetal93}
             and labeled with the stellar mass (in $M_\odot$). 
            }
   \label{fig:evolpehb}
\end{figure}

Because the time since its departure from the AGB is relatively long for \bd\ 
due to its low mass (Tab\@.~\ref{tab:mlrd}), all matter ejected previously on the AGB
(as well as the PN) may have dissipated into the ambient ISM. Thus, the
result of \citet{frew10} does not stringently exclude a post-AGB evolution.
However, an alternative stellar evolution may be possible.
\citet{bergeronetal94} discussed that the progenitors of most
DAO-type WDs are post-EHB stars ($M\,\approx\,0.48\,M_\odot$).
Recently, \citet{gianninas10} presented a spectral analysis of 29 DAO- and 18 hot DA-type WDs
and found that their DAOs (Fig.~\ref{fig:evolpagb}) are hotter and more massive that previously determined. 
Accordingly, they concluded that most of these (\emph{``a rather mixed bag of objects''}) are products 
of post-AGB evolution. 
\bd\ is located amongst the low-mass stars of their sample (Fig.~\ref{fig:evolpehb})
and consequently, a post-EHB evolution cannot be ruled out.
In this scenario, the stars evolve directly from the EHB to the WD state \citep[AGB-manqu\'e stars,][]{greggiorenzini90}.
\citet{bergeronetal94} proposed that a weak mass-loss may keep
enough helium in the line-forming regions \citep{unglaubbues98,unglaubbues00}
to exhibit strong \Ion{He}{2} absorption lines.
Figure~\ref{fig:evolpehb} shows \bd\ in comparison to post-EHB
evolutionary tracks \citep{dormanetal93}. One can estimate that an
$M\,\approx\,0.48\,M_\odot$ track (with solar helium content on the horizontal branch, HB) 
will match the position of \bd. (Dorman's tracks for masses higher than $> 0.475\,M_\odot$ and solar helium content
are not fully calculated down to lower \Teff\ and higher \logg.)
On its way from the EHB to its present position, i.e\@., towards higher \Teff\ and
higher \logg, gravitational settling, radiative levitation, and a weak wind are
interacting and, thus, the atmospheric abundances may deviate from the HB abundances.

\begin{acknowledgements}
MZ was supported by the German Research Foundation (DFG, grant WE1312/38-1).
TR is supported by the German Aerospace Center (DLR, grant 05\,OR\,0806).  
This research has made use of the SIMBAD database, operated at CDS, Strasbourg, France.
This research has made use of NASA's Astrophysics Data System.
This work used the profile-fitting procedure \emph{OWENS} developed by M\@. Lemoine and the \emph{FUSE} French Team.
Some of the data presented in this paper were obtained from the 
Mikulski Archive for Space Telescopes (MAST). STScI is operated by the 
Association of Universities for Research in Astronomy, Inc., under NASA 
contract NAS5-26555. Support for MAST for non-HST data is provided by 
the NASA Office of Space Science via grant NNX09AF08G and by other 
grants and contracts.
\end{acknowledgements}

\bibliographystyle{aa}
\bibliography{19536}

\end{document}